\documentclass[natbib]{svjour3}                     
\smartqed  
\usepackage{graphicx}
 \usepackage{mathptmx}      
\usepackage{aps-bibstyle}  

\usepackage{url}
\usepackage{mathptmx}       
\usepackage{helvet}         
\usepackage{courier}        
\usepackage{type1cm}        
\usepackage{makeidx}         
\usepackage{graphicx}        
\usepackage{multicol}        
\usepackage[bottom]{footmisc}
\usepackage[export]{adjustbox}
\usepackage{amsmath}
\usepackage{amssymb}

\makeindex             

 \journalname{Space Science Reviews}
\newcommand{\cha}{{\sl Chandra}}
\newcommand{\xmm}{XMM-{\sl Newton}}
\newcommand{\nust}{{\sl NuSTAR}}
\newcommand{\msh}{MSH 15$-$5{\sl 2}}

\catcode`\@=11
\newcommand{\gapprox}{\mathrel{\mathpalette\@versim>}}
\newcommand{\lapprox}{\mathrel{\mathpalette\@versim<}}
\newcommand{\propapprox}{\mathrel{\mathpalette\@versim\propto}}
\newcommand{\@versim}[2]
  {\lower3.1truept\vbox{\baselineskip0pt\lineskip0.5truept
\ialign{$\m@th#1\hfil##\hfil$\crcr#2\crcr\sim\crcr}}}
\catcode`\@=12
\newcommand{\farcm}{\mbox{\ensuremath{.\mkern-4mu^\prime}}}
\newcommand{\farcs}{\mbox{\ensuremath{.\!\!^{\prime\prime}}}}

\newcommand{\wmap}{{\it WMAP}}
\newcommand{\planck}{{\it Planck}}
\newcommand{\rosat}{{\it ROSAT}}

\newcommand{\chan}{{\it Chandra}}

\newcommand{\agile}{\emph{AGILE}}

\newcommand{\lat}{\emph{Fermi}-LAT}
\newcommand{\hess}{H.E.S.S.}

\newcommand{\gammaray}{$\gamma$-ray}

\newcommand{\xray}{X-ray}
\newcommand{\xrays}{X-rays}

\def\amin{\ifmmode^{\prime}\else$^{\prime}$\fi}
\def\asec{\ifmmode^{\prime\prime}\else$^{\prime\prime}$\fi}
\def\d{$^\circ$}

\def\eg{{\it e.g.}}

\def\ie{{\em i.e.}}

\begin{document}

\title{Pulsar-wind nebulae and magnetar outflows:  observations at
radio, X-ray, and gamma-ray wavelengths
}

\titlerunning{Observations of PWN Outflows}        

\author{Stephen P. Reynolds         \and
       George G. Pavlov \and
       Oleg Kargaltsev \and
       Noel Klingler \and
       Matthieu Renaud \and
       Sandro Mereghetti 
}


\institute{S. Reynolds \at
              Physics Department, North Carolina State University \\
              Tel.: +01-919-515-7751\\
              Fax: +01-919-515-6538\\
              \email{reynolds@ncsu.edu}           
}


\date{Received: date / Accepted: date}

\maketitle

\begin{abstract}
We review observations of several classes of neutron-star-powered
outflows: pulsar-wind nebulae (PWNe) inside shell supernova remnants
(SNRs), PWNe interacting directly with interstellar medium (ISM), and
magnetar-powered outflows.  We describe radio, X-ray, and gamma-ray
observations of PWNe, focusing first on integrated spectral-energy
distributions (SEDs) and global spectral properties.  High-resolution
X-ray imaging of PWNe shows a bewildering array of morphologies, with
jets, trails, and other structures.  Several of the 23 so far
identified magnetars show evidence for continuous or sporadic emission
of material, sometimes associated with giant flares, and a few
``magnetar-wind nebula'' have been recently identified.

\end{abstract}

\section{Introduction}
\label{intro}

\begin{figure}
\centerline{\includegraphics[width=3truein]{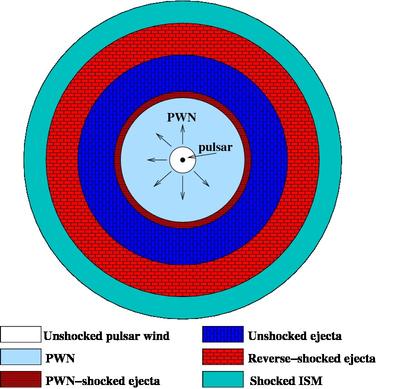}}
\caption{Cartoon of PWN inside a shell SNR.  The SNR blast wave is
the outermost circle.}
\label{fig:cartoon}
\end{figure}

Pulsars emit relativistic winds in a variety of forms, not well
understood at this time.  The outflows consist of some combination of
highly relativistic leptons ($e^+/e^-$ pairs) and perhaps ions as
well, and magnetic field.  The winds initially appear to be ``dark''
because they are cold in the comoving frame, but become thermalized
somehow at a wind termination shock, which may or may not resemble a
traditional perpendicular shock.  Its location is fixed by pressure
balance between the outgoing wind and the local ambient medium, which
is either a shell supernova-remnant (SNR) interior (see
Fig.~\ref{fig:cartoon}), or for older pulsars that have outlived their
SNR, undisturbed interstellar medium (ISM).  Beyond this point,
radiation from the outflows is apparent, and the observed object is
known as a pulsar-wind nebula (PWN).  [See \cite{gaensler06} and
\cite{kargaltsevpavlov2008} for general reviews of PWNe.]  The broadband
spatially integrated spectral-energy distribution (SED) appears to
consist of two parts: a lower-energy spectrum of synchrotron emission,
responsible for emission from low radio frequencies to X-rays and into
the MeV range (in the few cases where it is detectable at those
energies), and a higher-energy spectrum in the GeV-TeV range, in most
cases attributed to inverse-Compton upscattering of any of various
possible photon fields: the PWN's own synchrotron photons
(``synchrotron self-Compton,'' SSC), the cosmic microwave background
(ICCMB), or local optical/IR radiation fields.  In a few cases, it may
be that this emission is produced by relativistic ions colliding
inelastically with thermal gas and producing both charged and neutral
pions.  The $\pi^0$ mesons decay to photons, producing a continuum
above the kinematic threshhold of about 150 MeV.  But this hadronic
process is not thought to be a major contributor to most PWN GeV-TeV
spectra.

We structure this review mainly from young to older objects.  Pulsars
live far longer than their natal supernova remnant, so a relatively
small fraction of PWNe should still be found within a SNR.  However,
this subclass offers an excellent chance to understand pulsar
outflows, as information from the SNR can add to what can be deduced
from the PWN alone.  Other PWNe without clear evidence of a
surrounding shell still have unmistakable signs of youth.  The first
part of this review will focus on those two subclasses, on their
integrated SEDs and on their spatially resolved spectral properties.
Pulsars older than the typical lifetime of an SNR continue to emit
winds, which interact directly with the surrounding medium.  The
second part of this review deals with radio and X-ray emission from
these PWNe.  Some of both young and older X-ray PWNe emit gamma-rays
as well, but the gamma-ray class is predominantly made up of much
older objects which may be unprepossessing or undetectable at longer
wavelengths.  The third part of the review covers the gamma-ray
properties of PWNe of all ages.  Finally, magnetars have unique
characteristics in terms of outflows; these characteristics are
described in the last part of this chapter.

While the gross properties of PWNe (mean sizes, integrated spectra)
provide one kind of information on the basic physics, a great deal can
also be learned from detailed morphological investigations.  The {\sl
  Chandra} X-ray Observatory has provided a rich collection of imaging
information down to sub-arcsecond scales, and we review that
collection as well.  Many of these objects are also older, with the
pulsar wind interacting directly with ISM, sometimes in a well-defined
bow shock, but in other cases in unusual and perplexing morphologies.
Finally, we describe evidence for outflows from magnetars, neutron
stars with very strong magnetic fields and a propensity to emit giant
flares.  Evidence for ejection of material seems strong. While clear
cases of steady outflows (``magnetar-wind nebulae'') have not yet been
firmly confirmed, candidates have been identified.  We also mention
gamma-ray binaries, in which the relativistic pulsar wind interacts
with a wind from the binary companion, but 
these systems are described more fully elsewhere.

\section{Synchrotron emission from young pulsar-wind nebulae: radio to X-rays}
\label{sec:spr}

The synchrotron spectrum of PWNe contains the most specific
information about the particle spectrum injected into the nebula at
the wind termination shock (WTS), though the unknown magnetic-field
structure can complicate the extraction of that information.
Furthermore, the injected spectrum can change with time and location,
due to evolutionary effects and particle propagation, e.g., advection or
diffusion. Disentangling these effects is essential to understand
the nature of particle energization at the WTS.  

\begin{figure}
\centerline{\includegraphics[width=1.9truein,angle=270,origin=c]{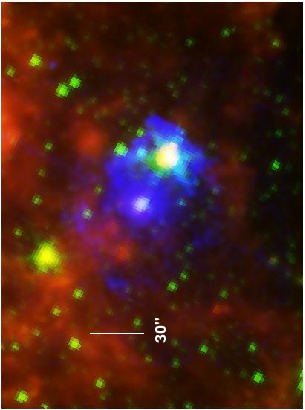}
    \hskip0.2truein
            \includegraphics[width=2.3truein]{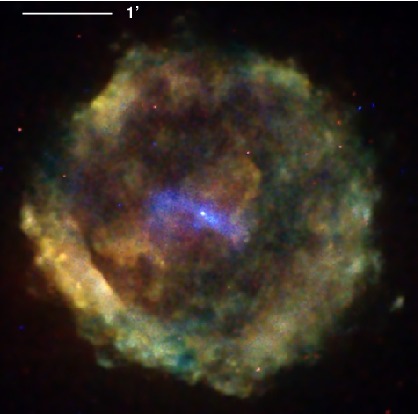}}
\centerline{\includegraphics[width=2.75truein]{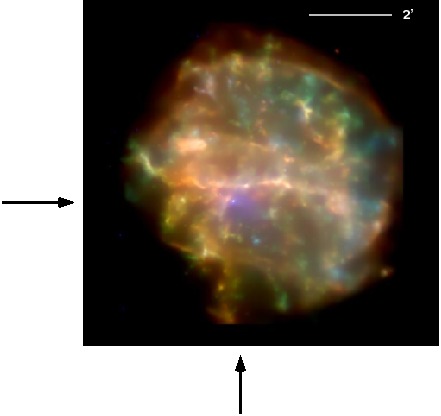}
    \hskip0.2truein
           \includegraphics[width=2.3truein]{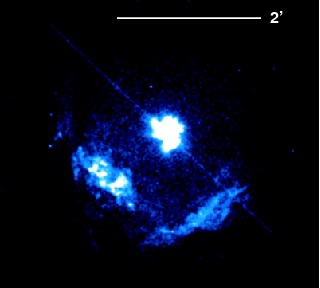}}
\caption{Upper left: B0540$-$693 in the Large Magellanic Cloud
  (Williams et al.~2008).  Red, {\sl Spitzer} 8 $\mu$m; green, {\sl
    Spitzer} 3.6 $\mu$m; blue, {\sl Chandra} 0.5 -- 8 keV. The pulsar
  and surrounding PWN is barely resolved at the center of the frame,
  just above and to the left of the bright foreground star.  Upper
  right: G11.2$-$0.3 with {\sl Chandra} (Borkowski et al.~2016). Red:
  0.6 -- 1.2 keV; green, 1.2 -- 3.3 keV; blue, 3.3 -- 8 keV.  The PWN
  is visible in blue.  Lower left: G292.0+1.8 with {\sl Chandra}
  \citep{park07}.  Red: 0.3 -- 0.8 keV; green, 0.8 -- 1.7 keV; blue,
  1.7 -- 8 keV.  The PWN is the purplish region slightly SE of the
  center, and the pulsar (indicated by arrows) is at its NE
  edge. Lower right: Kes 75 with {\sl Chandra} \citep{gavriil08}.  The
  pulsar is highly piled up, as the readout streak indicates.}
\label{fig:4pwne}
\end{figure}

Figure~\ref{fig:4pwne} illustrates four PWN/SNR combinations, in which
the pulsars are well studied.  The PWNe are asymmetric and
structured, and for G292.0+1.8, not centered on the pulsar.  However,
these systems are amenable to detailed interpretation based on 
analysis of both the PWN nonthermal radiation and the properties of
the surrounding SNR.  In both B0540$-$693 (Williams et al.~2008) and
G11.2$-$0.3 (Borkowski et al.~2016), the PWN's interaction with the
SNR inner ejecta is important in the overall characterization of the
object.  

Young objects like the Crab or 3C 58 (unlikely to be the
remnant of an event in 1181 AD, but still only of order 2000 yr old;
Chevalier 2005), without clear evidence of a shell, also show
properties consistent with expansion into a low-density medium which
is probably itself expanding.  Simple 1-D models show that as long as
the pulsar maintains its original energy-loss rate, the PWN expands
into uniformly expanding ejecta with radius $R \propto t^{6/5}$
\citep{rc84,vds01}, producing a shock wave into the inner ejecta that
strengthens with time.  (See Fig.~\ref{fig:cartoon}).

These two classes of young PWNe, with and without SNR shells, will be
the focus of this section.

\subsection{General properties of the PWN synchrotron spectrum}
\label{sec:genprops}

\begin{figure}
\includegraphics[width=4truein]{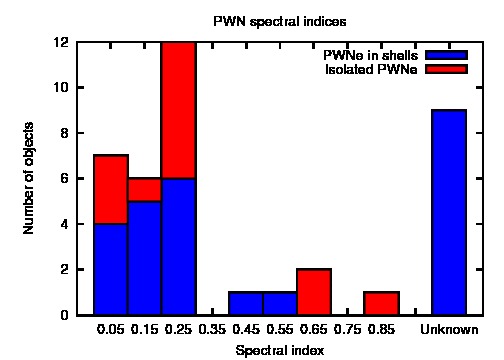}
\caption{Distribution of radio spectral indices $\alpha$ ($S_\nu \propto 
\nu^{-\alpha}$) of 39 PWNe in shells
(blue) and isolated (red).  Many of the former have poorly known
radio spectral indices.  Typical uncertainties on $\alpha$ are $\pm 0.1$.
The three steep-spectrum PWNe ($\alpha \ge 0.6$) are all anomalous in
other ways as well.}
\label{fig:spixhist}
\end{figure}

Pulsar-wind nebulae were originally defined by radio properties:
``flat'' radio spectrum (that is, energy spectral index $\alpha \sim 0
- 0.4$ where $S_\nu \propto \nu^{-\alpha}$ is the energy flux),
center-brightened morphology, and high radio polarization
\citep[e.g.,][]{weiler78}.  The catalog remained small, however, until
the high spatial resolution of \cha\ and \xmm\ allowed the
identification of PWNe in shells and in confused regions, at which
point a large increase in identified PWNe began which continues to
this day.  The catalog by \cite{ref:kargaltsev13} lists 70 X-ray PWNe with
known pulsars, and 6 more objects identified at gamma-ray energies
without currently known X-ray counterparts.  Many of the PWNe
discovered at X-ray or gamma-ray energies have weak, poorly known, or
undetectable radio emission.  In Green's catalog of SNRs and PWNe
\citep{green09}, 26 PWNe in shells (formerly known as ``composite''
SNRs) and another 13 isolated PWNe (once called ``plerions'') are
listed.  This catalog began as a listing of radio SNRs, and does not
include pulsar bow-shock nebulae or other manifestations of pulsar
outflows without clear radio counterparts.  With the addition of two
more recent discoveries, Fig.~\ref{fig:spixhist} shows the
distribution of PWN radio spectral indices.  Since there are a few good
cases of steeper-spectrum PWNe ($\alpha > 0.4$), it is possible that
more such objects exist but have been selected against.  However, the
three steep-spectrum PWNe ($\alpha \ge 0.6$) are all anomalous in
other ways as well.  It is also possible that some very flat-spectrum
objects ($\alpha \sim 0.1$) have been mistaken for H II regions, since
strong radio polarization, the usual discriminator, is not found in
all objects.

\begin{figure}
 \centerline{ 
  \includegraphics[width=2.5truein]{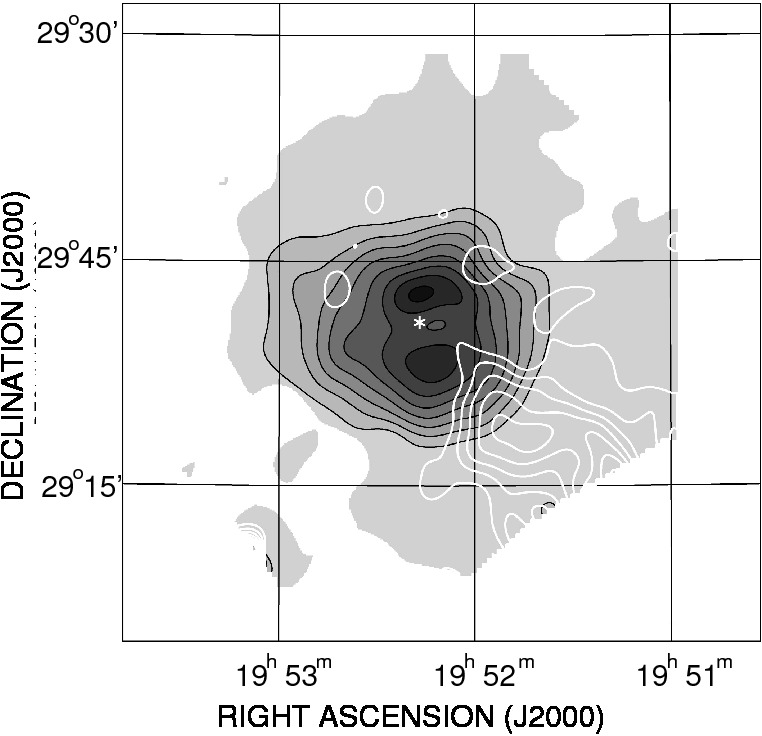} \hskip0.1truein
  \includegraphics[width=2.5truein]{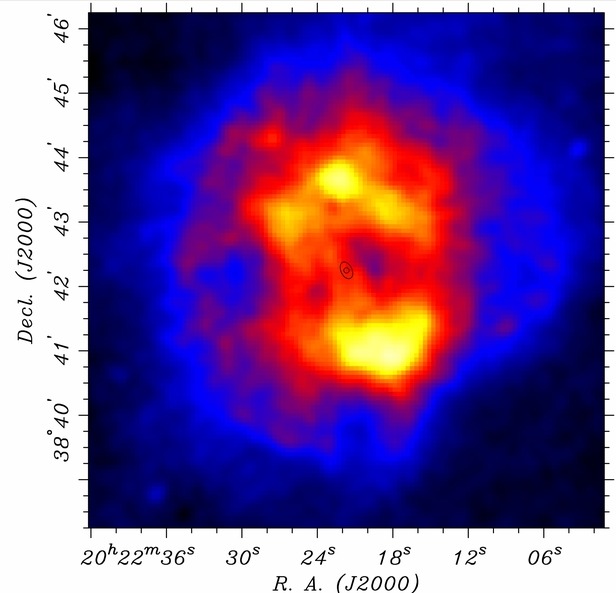} \hskip0.1truein}
\centerline{
  \includegraphics[width=3truein]{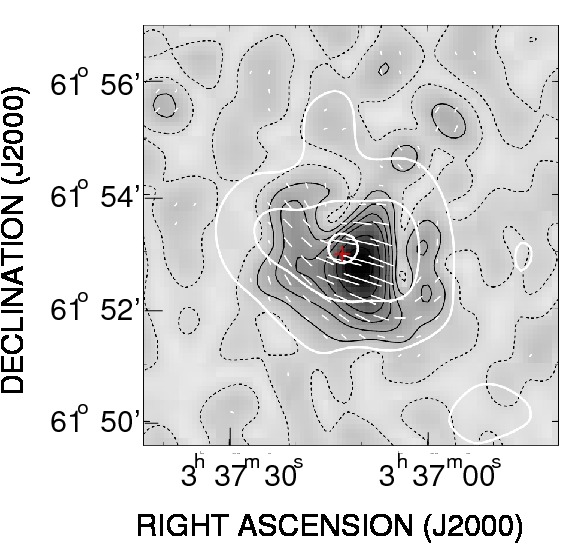} }
\caption{Three PWNe with steep radio spectra.  Upper left: DA 495
  (G65.7+1.2) at 1.4 GHz.  White contours: a background H II region.
  Asterisk: a compact X-ray source \citep{kothes08}. Upper right: G76.9+1.0
  \citep{arzoumanian11} at 1.4 GHz.  Almost invisible contours of extent
about $40''$ near the center show the X-ray nebula.  Bottom:
G141.2+5.0 at 1.4 GHz in polarized intensity (greyscale) with contours
indicating total intensity \citep{kothes14}, and the red cross indicating
a pointlike X-ray source \citep{reynolds16}.}
\label{fig:3steep}
\end{figure}

The three anomalous steep-spectrum PWNe are shown in
Fig.~\ref{fig:3steep}.  The first two, DA 495 and G76.9+1.0, with
$\alpha = 0.87 \pm 0.1$ \citep{kothes08} and $0.62 \pm 0.04$
\citep{landecker93}, have strikingly similar double-lobed morphology,
with tiny extended X-ray sources in the center of each.  G141.2+5.0
\citep{reynolds16} shows a simple center-brightened morphology, but
with a central X-ray source unresolved by \cha. It is not known
how these objects fit into the overall PWN scheme; no data are
available for any at other wavelengths.

\begin{figure}
\centerline{\includegraphics[width=3truein]{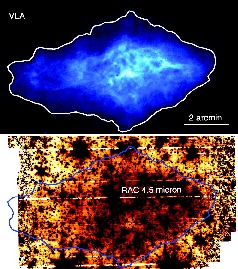}}
\vskip3mm
\centerline{\includegraphics[width=2truein]{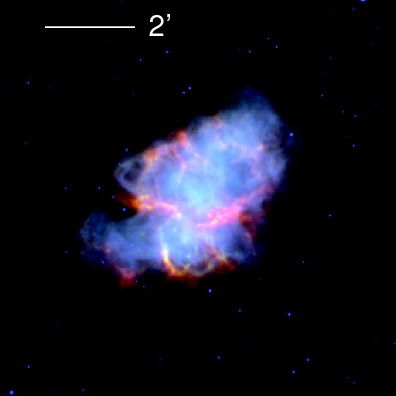}
    \hskip2mm
\includegraphics[width=2.5truein]{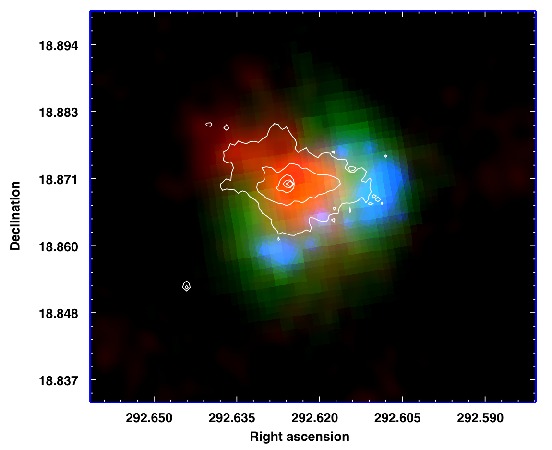}}
\vskip3mm
 \centerline{\includegraphics[width=5truein]{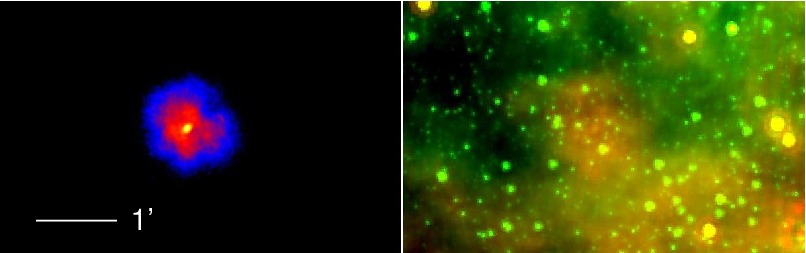}}
\caption{Top: 3C 58 \citep{slane08}.  (Upper: radio, VLA at 1.4 GHz
  \citep{reynolds88}; lower, IR, {\sl Spitzer} at 4.5 $\mu$m.)  Middle
  left: Crab Nebula \citep{temim06}. Red, 24 $\mu$m (mainly [O IV] at
  26 $\mu$m); green, 8 $\mu$m (mainly [Ar II] at 7 $\mu$m); blue, 3.6
  $\mu$m (mainly synchrotron continuum).  Middle right: G54.1+0.3
  \citep{temim10}.  Contours, X-ray.  Red, radio.  Green, 70 $\mu$m.
  Blue, 24 $\mu$m.  Lower: G21.5-0.9 \citep{zajczyk12}.  Left:
  \cha\ image.  Right: Red, 24 $\mu$m; green, 8 $\mu$m.}
\label{fig:spitzerims}
\end{figure}

For many PWNe, no other observations are available at frequencies
below the X-ray regime.  But a few have been imaged with {\sl Spitzer}
(see Fig.~\ref{fig:spitzerims} and Table~\ref{tab:spitzer}).  Much of
the emission is thermal radiation from dust grains, or fine-structure
spectral lines, but some synchrotron continuum is evident in the Crab,
3C 58, and G21.5$-$0.9.

\begin{table}
\caption{{\sl Spitzer} observations of PWNe}
\label{tab:spitzer}       
\begin{tabular}{lllll}
\hline\noalign{\smallskip}
Object & Type & Detectors$^a$ & Result & Reference\\
\noalign{\smallskip}\hline\noalign{\smallskip}
Crab  &PWN  &I, M, S &Images, spectrum &\cite{temim06}\\
G21.5--0.9  &PWN           &I, M    &Images& \cite{zajczyk12}\\
3C58       &PWN           &I, M     &Images & \cite{slane08}\\
G54.1+0.3  &PWN, IR shell &I, M, S &Images, spectrum & \cite{temim10}\\
B0540--693  &PWN, shell    &I, M, S &Images, spectrum& \cite{williams08}\\
\noalign{\smallskip}\hline\noalign{\smallskip}
\multicolumn{5}{l}
  {$^a$I: IRAC (3.6, 4.5, 5.8, \& 8.0 $\mu$m). M: MIPS (24, 70, 160 $\mu$m).
S: IR Spectrograph (5 -- 35 $\mu$m)}\\
\end{tabular}
\end{table}

\begin{figure}
  \centerline{\includegraphics[width=2truein]{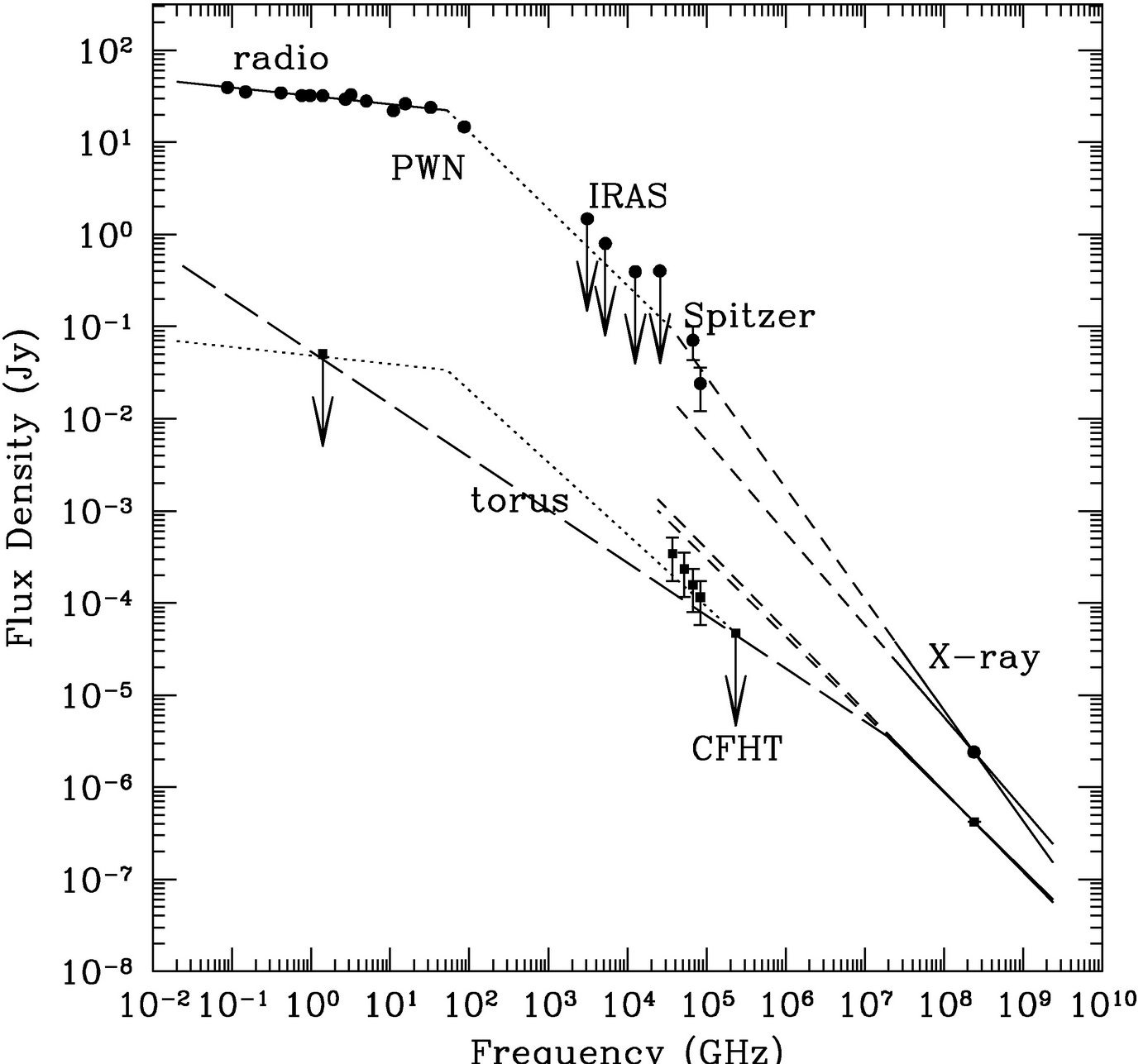} \hskip5mm
      \includegraphics[width=2.2truein]{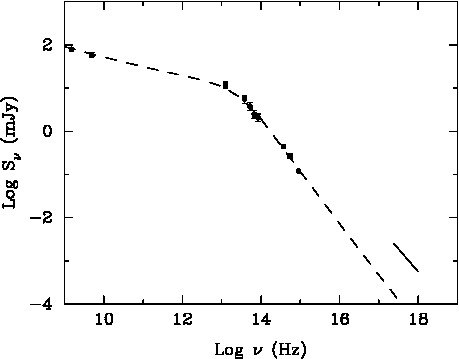}}
\caption{Left: SED of 3C 58 \citep{slane08}.  The spatially integrated
  spectrum is above, and is roughly describable by two power-laws
  of slopes $\alpha$ of 0.1 and $\sim 1$, with
  a break near 100 GHz.  (The small-scale torus requires a
  considerably more complex spectrum.) Right: SED of B0540$-$693 in
  the Large Magellanic Cloud \citep{williams08}.  Triangles are {\sl
    Spitzer} observations.  The solid line is the \cha\ spectrum
  \citep{kaaret01}, which is affected by pileup in the pulsar.  The
  PWN X-ray spectrum does not appear to be consistent with the
  power-law extrapolation from the optical-IR \citep{serafimovich04}.}
\label{fig:spitzersed}
\end{figure}

\begin{figure}
\centerline{\includegraphics[width=3truein]{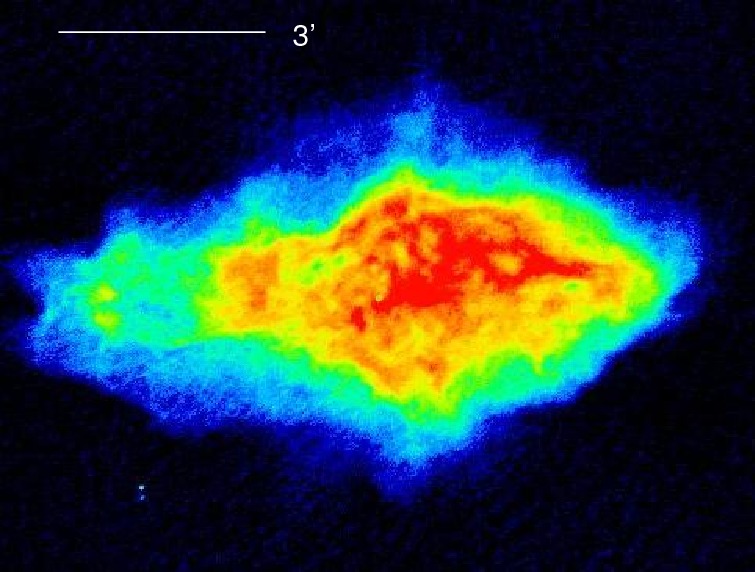}}
\centerline{\includegraphics[width=3truein]{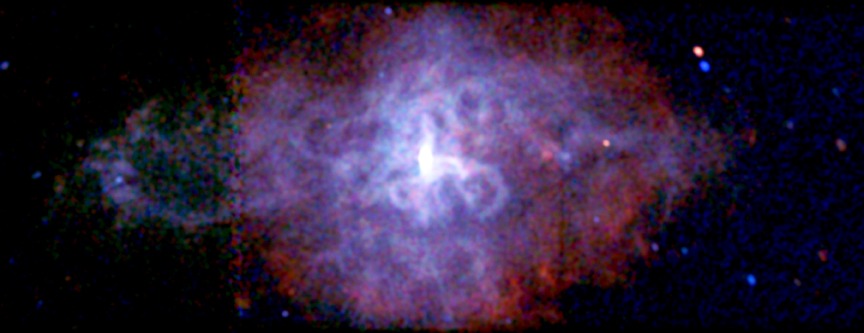}}
\caption{Top:  VLA image of 3C 58 at 1.4 GHz \citep{reynolds88}.
Bottom:  \cha\ X-rays \citep{slane08}.  Note the close correspondence
between detailed features, and the similar maximum extent in
radio and X-rays.}
\label{fig:3c58rx}
\end{figure}

\begin{figure}
\centerline{\includegraphics[width=2.4truein]{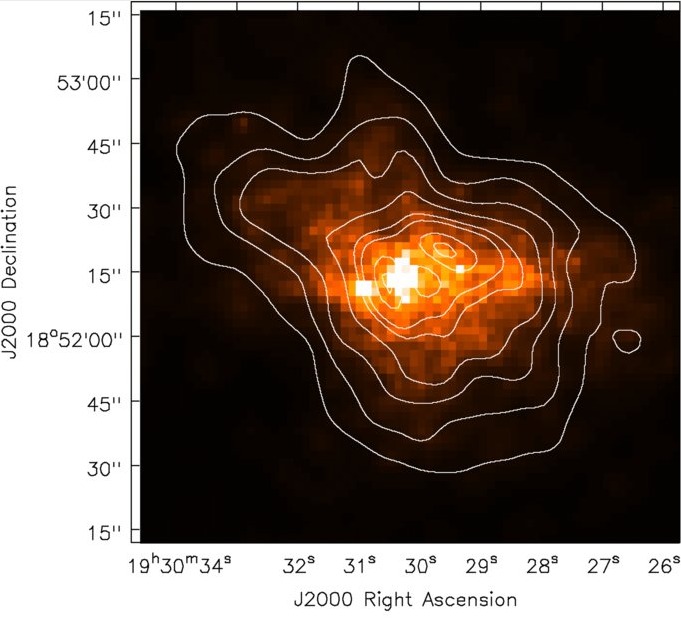}\hskip2mm
   \includegraphics[width=2.4truein]{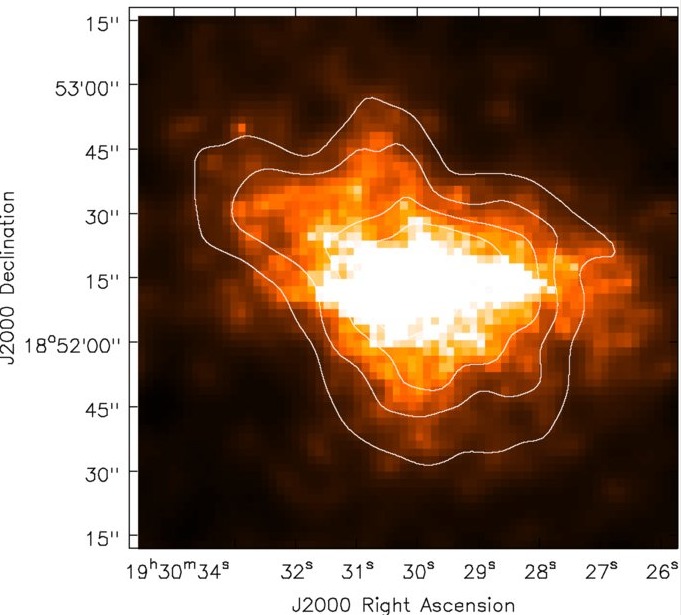}}
\caption{Images of G54.1+0.3 in radio (contours; VLA at 4.7 GHz) and 
X-ray (\cha), with two different stretches to emphasize the extent
of faint X-ray emission \citep{lang10}.}
\label{fig:g54rx}
\end{figure}

PWNe are often observed to be smaller at X-ray than at longer
wavelengths, as is clearly the case with the Crab Nebula.  However,
several other prominent PWNe show X-rays extending to the edges of the
radio contours, though with greater center-to-edge brightness
contrast.  Figures~\ref{fig:3c58rx} and~\ref{fig:g54rx} show that both
3C 58 and G54.1+0.3 have X-ray extents comparable to their radio
extents.

\subsection{Imaging PWNe above 10 keV with {\sl NuSTAR}}

\begin{figure}
\centerline{\includegraphics[width=3truein]{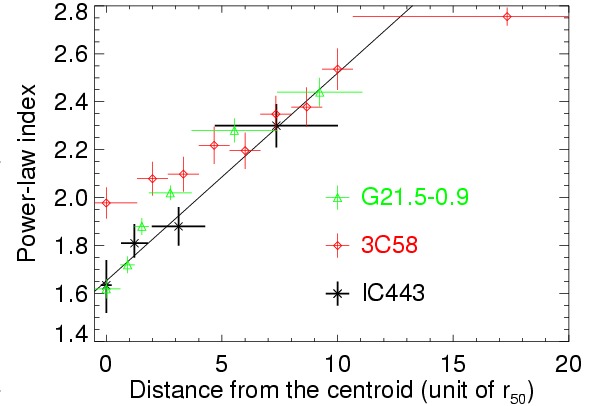}}
\caption{Steepening of the X-ray spectrum with scaled distance from the
center, for three PWNe \citep{bocchino01}.  The distance is in units of
the distance at which the surface brightness has dropped a factor of 2
below its peak.  The line is a straight-line fit to the IC 443 data
only.}
\label{fig:3gams}
\end{figure}

Spectral inhomogeneity in PWNe is commonly observed at X-ray energies,
where energy losses are becoming important.  Spectra steepen with
distance from the central pulsar, presumably as higher-energy
electrons are depleted.  The steepening trend tends to begin
immediately, as shown in Fig.~\ref{fig:3gams} for three PWNe
\citep{bocchino01}.  This behavior is explored more fully below.  This
means, however, that integrated SEDs may hide important spectral
variations. Unfortunately, until recently true imaging at X-ray
energies above about 10 keV was not available.  However, the
\nust\ mission, launched in 2012 \citep{harrison13}, has brought that
capability to the study of PWNe.  Results of observations of three
PWNe with \nust\ are summarized below; details are in the primary
publications listed.

\subsubsection{Crab Nebula}

\begin{figure}
\centerline{\includegraphics[width=4truein]{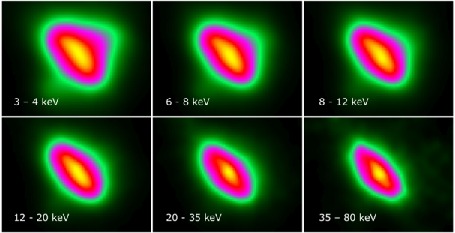}}
\caption{The Crab Nebula with \nust\ \citep{madsen15b}, after
  maximum-entropy deconvolution. The nebula shrinks with increasing
  photon energy, at different rates in different directions.}
\label{fig:crabnustar}
\end{figure}

\begin{figure}
  \centerline{\includegraphics[width=2truein]{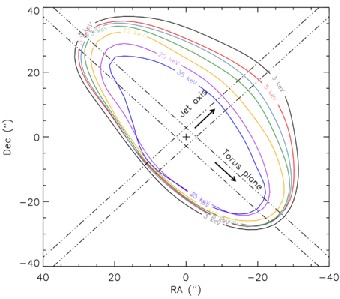}
  \hskip3mm
  \includegraphics[width=2.5truein]{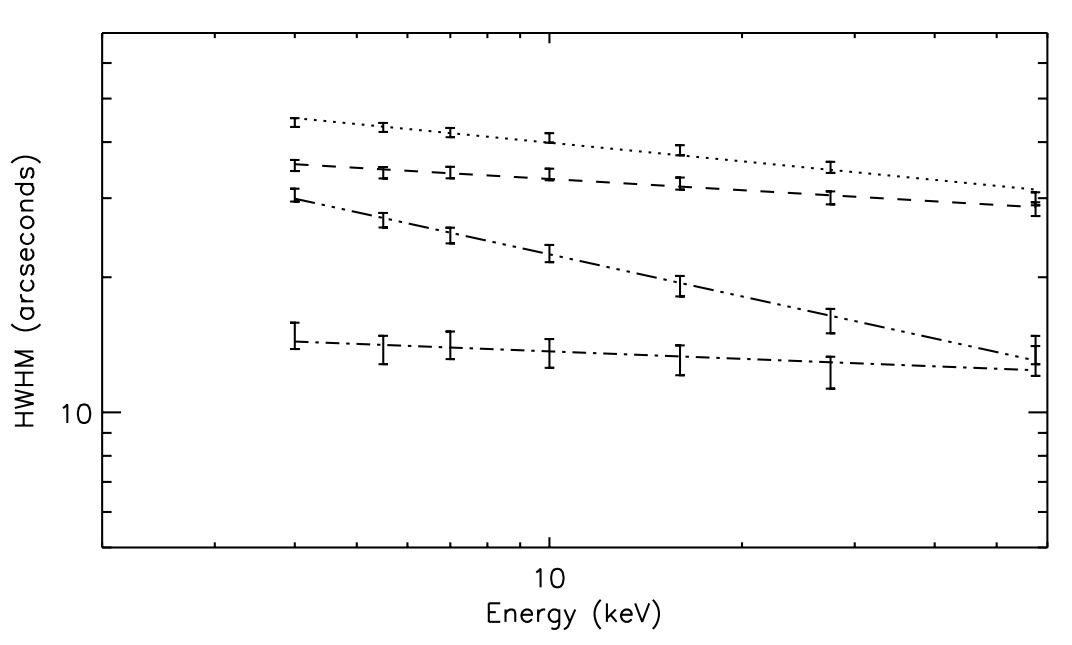}}
\caption{Left: Crab 50\% peak intensity contours at different energies
  \citep{madsen15b}.  From outside in, 3 -- 5, 5 -- 6, 6 -- 8, 8 --
  12, 12 -- 20, 20 --35, and 35 -- 78 keV.  Right: Fits to the HWHM
  along different directions.  From top to bottom: NE torus plane, SW
  torus plane, NW jet axis, and SE jet axis.}
\label{fig:crabshrink}
\end{figure}

\begin{figure}
  \centerline{\includegraphics[width=2.6truein]{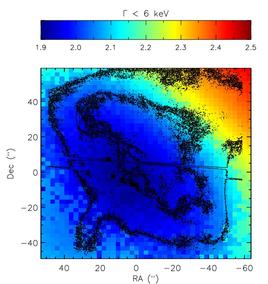}
  \hskip2mm
  \includegraphics[width=2.7truein]{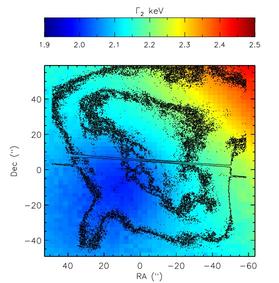}}
\caption{Left:  Crab photon index $\Gamma$ below 6 keV;
right:  $\Gamma$ above 10 keV \citep{madsen15b}.  Contours are
from \cha\ \citep{mori04}.}
\label{fig:crabgams}
\end{figure}

For PWNe with a bright pulsar, disentangling the PWN emission from the
pulsar, and then searching for spectral variations, requires good
resolution in both time and space.  \cite{madsen15b} describe this
process in detail for the Crab.  The observations were divided into 13
phase intervals of the pulsar rotation period, and only bins 10 -- 12,
where pulsar emission was negligible, were used for the PWN
spatial/spectral analysis (see Figs.~\ref{fig:crabnustar}
and~\ref{fig:crabshrink}). The spatially and temporally integrated
signal was calibrated to the accepted power-law shape with $\Gamma =
2.1$ [$F(E) \propto E^{-\Gamma}$ photons cm$^{-2}$ s$^{-1}$
keV$^{-1}$].  But the spectrum is substantially harder at smaller
radii.  This was known previously through \cha\ observations
\citep{mori04}. However, \nust\ has revealed another important
feature: the spectrum of the inner nebula (primarily the torus)
appears to steepen by a substantial amount, $\Delta \Gamma \cong
0.25$, above a break energy of about 9 keV.  Fig.~\ref{fig:crabgams}
shows the low and high-energy $\Gamma$ values, based on the
\nust\ data.

The original spherical, steady-state MHD model of
\cite{kennel84a,kennel84b} predicted how the size of the Crab Nebula
in the X-ray range should scale with photon energy $E$: $R \propto
E^m$ with $m = -1/9$.  This prediction was consistent with 
results found by several
sounding-rocket measurements with lunar occultations in the 1970's
\citep{kestenbaum75,ku76} which obtained ${\rm FWHM} \propto
E^{-0.148 \pm 0.012}$, fitting Gaussians to the data.  This result was
regarded as adequate agreement.  The \nust\ data show that the
shrinkage is spatially varying (Figs.~\ref{fig:crabnustar}
and~\ref{fig:crabshrink}).  The fitted values of $m$ are $-0.086 \pm
0.025$ along the NE direction and $-0.073 \pm 0.028$ along the SW, or
about $-0.08 \pm 0.03$ along the torus plane.  For the SE, $m$ is
consistent with zero, but along the NW (``counter-jet'' direction),
the rate is substantially larger: $m = -0.218 \pm 0.040$.  The torus
rate is consistent with the \cite{kennel84a,kennel84b} prediction, but
the counter-jet clearly shrinks much more rapidly.

\subsubsection{G21.5$-$0.9}

This bright PWN was used for calibration by \nust\ and other missions.
It has been reported as a TeV source \citep[][see
  Fig.~\ref{fig:g21sedsize}]{ref:djannati-atai08}; while those data appear to
lie roughly on an extrapolation from the measurements of
\nust\ and INTEGRAL, the TeV emission must be due to a separate
process as normal synchrotron emission is limited to photon energies
below a few hudred MeV.  However, the radio -- X-ray spectrum appears
to be reasonably well described by two power-laws, requiring a
steepening in mid-IR by $\Delta \equiv \alpha(\mathrm{high}) -
\alpha(\mathrm{low}) \sim 1.0$, where $\alpha(\mathrm{high})$ is the
X-ray energy index ($\alpha_X \equiv \Gamma - 1$).   

\begin{figure}
\centerline{\includegraphics[width=5truein]{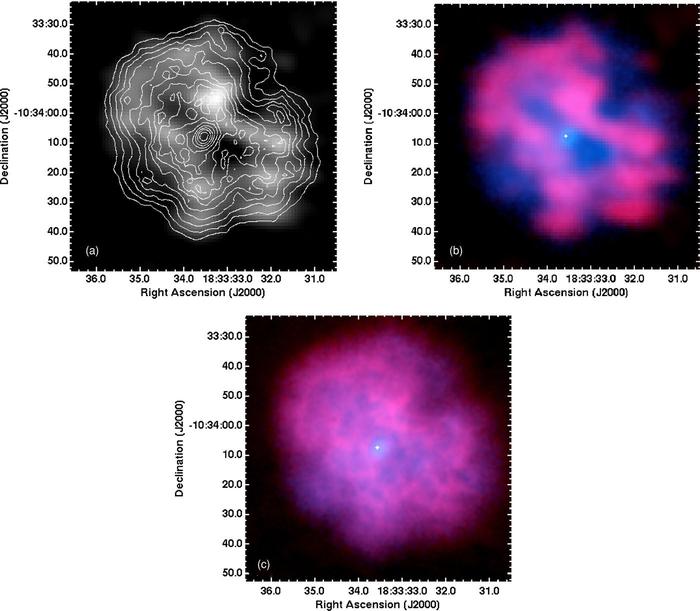}}
\vskip3truemm
\centerline{\includegraphics[width=3truein]{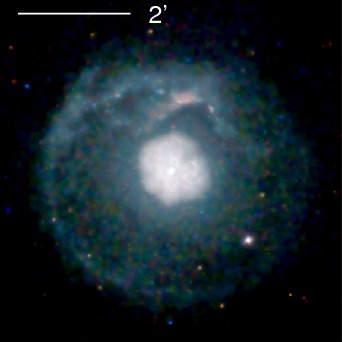}}
\caption{G21.5-0.9 in radio and X-rays \citep{matheson10}. Upper left: radio 
(greyscale; 22.3 GHz from the Nobeyama Millimeter-Wave
Array, resolution $8''$ \citep{fuerst88} 
and X-rays \citep[\cha\ contours;][]{matheson10}.  Upper right:  
22.3 GHz data in red, \cha\ in blue. Center:
4.75 GHz radio \citep[VLA;][]{bietenholz08} in red, same X-rays in blue.
Bottom: \cha\ \citep{matheson10}.}
\label{fig:g21rx}
\end{figure}

The fairly symmetric appearance of G21.5$-$0.9, and its similar size
and morphology in radio, IR, and X-rays, are unusual for PWNe.
Fig.~\ref{fig:g21rx} shows that the PWN sits in the center of an
apparent shell of diffuse X-rays.  This diffuse emission shows some
apparent limb-brightening in the SE and irregular structures to the N
which have thermal spectra and appear to be the SNR shell
\citep{bocchino05}, but there is also a substantial X-ray halo due to
scattering by interstellar dust providing the symmetric component of
diffuse emission that drops with radius \citep{bandiera04}.

\begin{figure}
\centerline{\includegraphics[width=5truein]{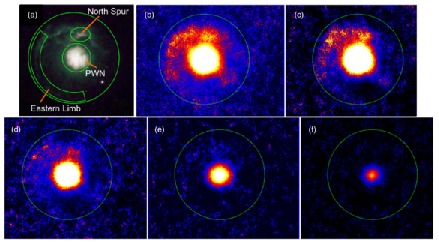}}
\caption{(a): \cha\ image of G21.5$-$0.9 between 3 and 6 keV
  \citep{matheson05}.  (b) -- (f): Deconvolved \nust\ images in
  various bands: (b) 3 -- 6 keV; (c) 6 -- 10 keV; (d) 10 -- 15 keV;
  (e) 15 -- 20 keV; and (f) 20 -- 25 keV. The green circles in panels
  (b) through (f) have radii of $165''$.}
\label{fig:g21nustarims}
\end{figure}

\nust\ observations \citep{nynka14} show substantial shrinkage with
X-ray energy (Fig.~\ref{fig:g21nustarims}, though the FWHM shrinkage
is less dramatic than the drop in total flux).
Fig.~\ref{fig:g21sedsize} shows the FWHM shrinkage, well described by
a power-law with $m = -0.21 \pm 0.01$, similar to the value found for
the Crab counter-jet.  The northern part of the shell is detectable in
the images to at least 20 keV, suggesting that part of the emission
may be nonthermal.  The \nust\ spectrum shows, as for the Crab torus,
a spectral softening that can be fit with a broken power-law, with
$\Gamma$ steepening from $1.996 \pm 0.013$ below a break energy $E_b =
9.7^{+1.2}_{-1.4}$ keV to $2.093 \pm 0.013$ above that energy.  (Dust
scattering efficiency drops with increasing photon energy and should
not affect the spectrum in the \nust\ energy range above 3 keV.)

\begin{figure}
\centerline{\includegraphics[width=2.5truein]{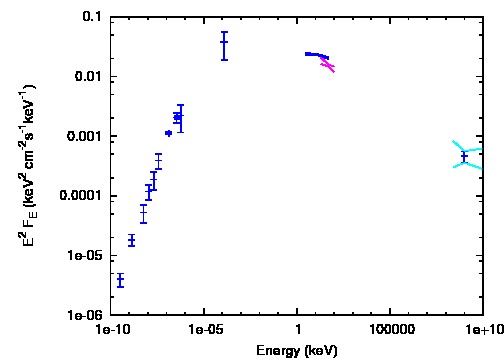} \hskip3mm
  \includegraphics[width=2.5truein]{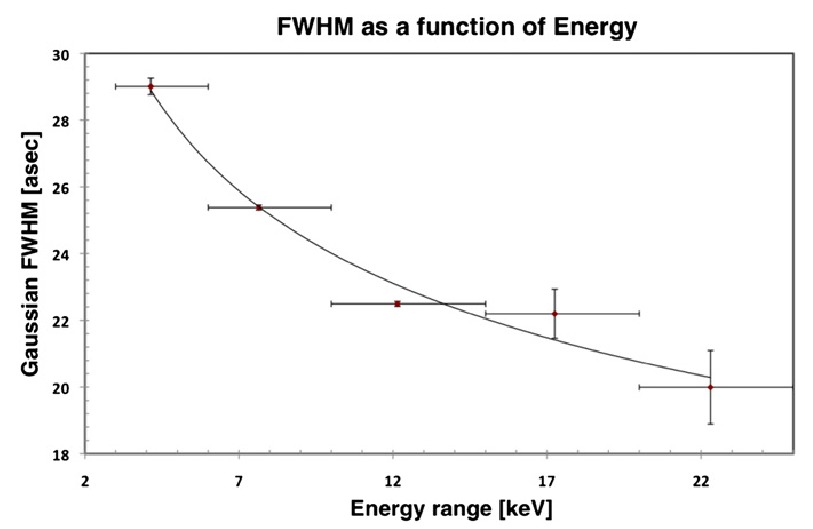}}
\caption{Left: G21.5$-$0.9 SED (PWN only).  Radio -- mm points from
  \cite{salter89}; IR point from \cite{zajczyk12}; blue bowtie,
  \nust\ \citep{nynka14}; magenta bowtie, INTEGRAL \citep{derosa09};
  TeV bowtie in cyan, H.E.S.S.  \citep{ref:djannati-atai08}. Right:
  Shrinkage of G21.5$-$0.9 with photon energy \citep{nynka14}.}
\label{fig:g21sedsize}
\end{figure}

\subsubsection{MSH 15$-$52}

\begin{figure}
\centerline{\includegraphics[width=5truein]{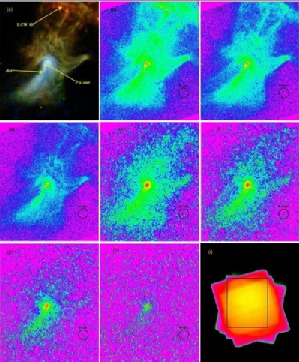}}
\caption{\msh\ with \cha\ and \nust\ \citep{an14}. From left to right, top to bottom:  \cha+\nust, 0.5 -- 40 keV; \cha, 0.5 -- 2 keV; \cha, 2 -- 4
  keV; \cha, 4 -- 7 keV; \nust, 3 -- 7 keV; \nust, 7 -- 12 keV; \nust,
  12 -- 25 keV; \nust, 25 -- 40 keV; \nust\ exposure map.  All frames
  except the last are $10' \times 12'$.}
\label{fig:mshims}
\end{figure}

\begin{figure}
\centerline{\includegraphics[width=2truein]{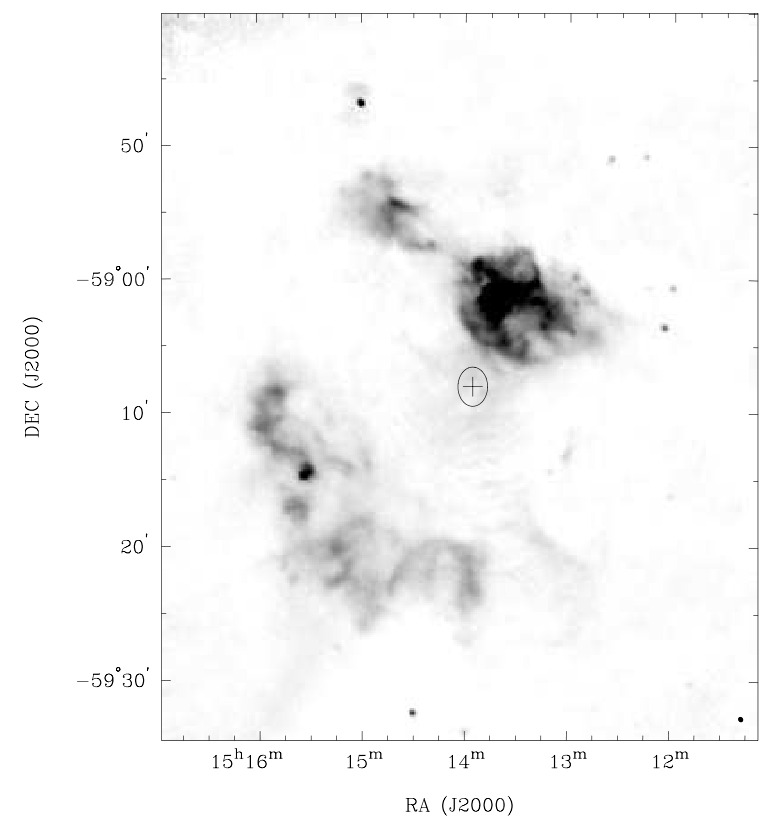}
\hskip3mm \includegraphics[width=3truein]{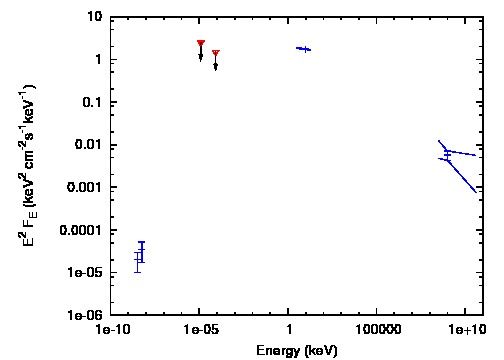}}
\caption{Left: Radio image of \msh\ at 1.3 GHz with ATCA
  \citep{gaensler99}.  The bright horseshoe to the N is RCW 89.
There is little evidence of the PWN; the cross marks the pulsar
location. Right:  SED for \msh.  Radio: \cite{gaensler02}.
IR upper limits are from \cite{koo11}.  X-rays: \cite{an14}.
TeV: \cite{aharonian05}.}
\label{fig:mshsed}
\end{figure}

This complex object has almost no radio counterpart
\citep{gaensler02}, but is bright at X-ray wavelengths
(Fig.~\ref{fig:mshims}).  It contains a well-known 150 ms pulsar, PSR
B1509-58, with a 1600-yr spindown timescale \citep{seward82}.  A
bright, curved jet extends to the SE, while long, straighter
``fingers'' reach to the NW to an H II region, RCW89, which contains
small knots of radio and X-ray emission.  A TeV detection has been
reported \citep{aharonian05}.  The full SED is shown in
Fig.~\ref{fig:mshsed}.

\begin{figure}
\centerline{\includegraphics[width=5truein]{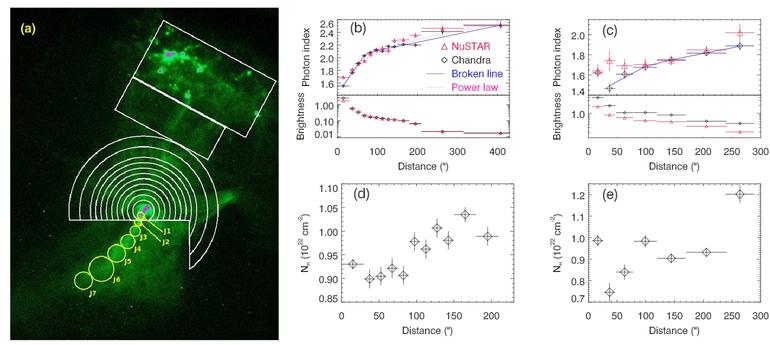}}
\caption{Left (panel (a)): Regions of \msh\ used for spatially
  resolved spectroscopy.  Panel (b): Photon index as a function of
  distance to the NW (in the partial annuli).  Panel (c): Photon index
  along the jet to the SE. Panels (d) and (e) give fitted column
  densities $N_H$ from {\sl Chandra} data as a function of distance
  from the pulsar, for the northern nebula and the jet, respectively.
  Figures from \cite{an14}.}
\label{fig:mshgamma}
\end{figure}

\begin{figure}
\centerline{\includegraphics[width=5truein]{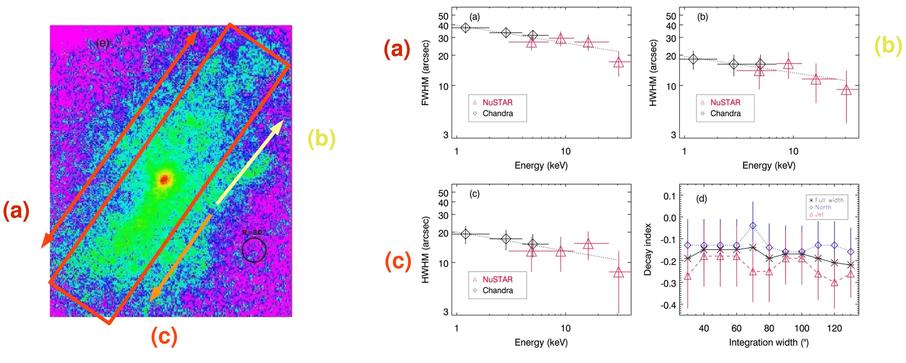}}
\caption{Left: Box used to extract profiles from \nust\ image of \msh.
  Profiles were summed along the minor dimension ($100''$), and
  extents as a function of energy displayed in panels to the right.
  Panel (a): FWHM of profiles centered on the pulsar extending in both
  directions. (b): Extending from the pulsar to the NW. (c) Extending
  from the pulsar to the SE (along the jet). (d) Energy shrinkage
  (``decay'') indices $m$ (with extents $\propto E^{m}$).  Plots from
  \cite{an14}.}
\label{fig:mshsize}
\end{figure}

\nust\ observations \citep{an14} show the same progressive
steepening evident in other PWNe.  Combined fitting with
\cha\ and \nust\ gives a rate that appears to slow with distance,
and can be described by two power-laws (Fig.~\ref{fig:mshgamma}).
This steepening is reflected in the broader measure of source extent
as a function of photon energy (Fig.~\ref{fig:mshsize}).  Profiles were
created by summing transversely in a rectangular box of width $100''$
extended along the jet axis.  Again, combined data from \cha\ and
\nust\ were used.  Power-laws describe the data fairly well, with $m
\sim -0.2$ (Fig.~\ref{fig:mshsize}), a value similar to that found for
G21.5$-$0.9 and for the counterjet of the Crab.

\subsection{Spectral breaks in PWNe}

The integrated synchrotron spectrum of all known PWNe steepens sharply
between radio and X-rays.  Steepening due to radiative losses is
expected, but steepening can also be caused by evolutionary effects,
such as the dropoff of pulsar luminosity after a few hundred years
\citep[e.g.,][]{pacini73}, or by intrinsic structure in the spectrum
of particles injected into the nebula at the wind termination shock.
It is important to be able to characterize the injected spectrum to
understand the physics of particle acceleration, so determining the
cause of spectral structure is a necessary part of using PWNe as
laboratories for the study of astrophysical particle energization.

\begin{figure}
\centerline{\includegraphics[width=2.5truein]{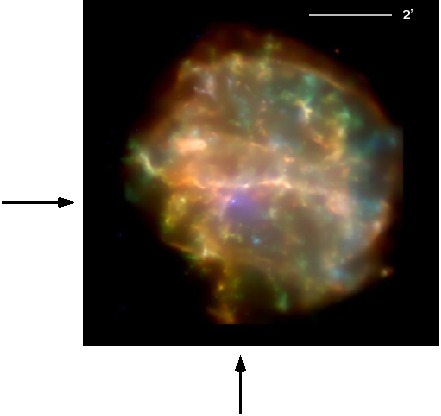}\hskip0.2truein
\includegraphics[width=2.5truein]{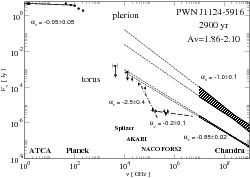}}
\caption{Left: The shell SNR G292.0+1.8, with the pulsar indicated and
  the PWN in blue \citep{park07}.  Right: SEDs of the entire PWN and
  of the torus [seen in VLT images at $H$ and $K_s$ bands;
  \cite{zharikov13}].  The torus spectrum is complex, but the PWN
  could be described by a double power-law with a single spectral
  break around 100 GHz.  However, the steepening $\Delta$ is almost
  1.}
\label{fig:g292sed}
\end{figure}

\begin{table}
\caption{Spectral breaks in PWNe}
\label{tab:breaks}
\begin{tabular}{lllll}
\hline\noalign{\smallskip}
Object   & $\alpha_{\rm radio}$  & $\alpha_{\rm X-ray}$  & $\Delta$ &Reference\\
\noalign{\smallskip}\hline\noalign{\smallskip}
Crab         &  0.3      &1.1   &0.8  & \cite{manchester93}\\
G11.2$-$0.3  &  $\sim 0$ &0.9   &0.9  & \cite{roberts03}  \\
Kes 75       &  0.2      &0.9   &0.7  & \cite{blanton96}  \\
G54.1+0.3    &0.28       &0.8   &0.5  & \cite{lu02} \\
3C 58        &0.1        &1.1   &1.0  & \cite{green92}\\
B0540$-$693  &0.25       &1.2   &0.95 & \cite{manchester93}\\
G21.5$-$0.9  &$\sim 0$   &1.0   &1.0  & \cite{salter89}, \cite{nynka14}\\
MSH 15$-$5{\it 2} & 0.2  &1.1   &0.9  & \cite{gaensler02}\\
G292.0+1.8   &0.05       &0.9   &0.85 & \cite{gaensler03}\\
\noalign{\smallskip}\hline\noalign{\smallskip}
\end{tabular}
\end{table}

While the small-scale torus of a PWN often has a complex,
non-monotonic SED (see, for example, Figs.~\ref{fig:spitzersed}
and~\ref{fig:g292sed} for the cases of 3C 58 and G292.0+1.8), the
larger-scale PWN often has simpler spectral behavior.  The spectrum of
an initially straight power-law distribution of electrons subject to
synchrotron losses steepens at an energy at which the source age
equals the synchrotron-loss time \citep{kardashev62}.  The amount of
spectral steepening $\Delta \equiv \alpha(\mathrm{high}) -
\alpha(\mathrm {low})$ is 0.5 in the case of a uniform source with
constant particle injection, but for inhomogeneous sources, naturally
arising from outflow, $\Delta$ can differ from 0.5
\citep{kennel84b,reynolds09}.  Since $\Delta > 0.5 $ is far more
common than $\Delta = 0.5$ (see Table~\ref{tab:breaks}, and
Fig.~\ref{fig:spitzersed} for the examples of 3C58 and 0540$-$693,
both with $\Delta \sim 1$), either these breaks are not due to losses
(in which case the absence of loss breaks poses serious modeling
problems), or PWNe are significantly inhomogeneous.  Outflow models,
however, predict spatial dependence of steepening which is at odds
with observations (uniform spectra until sudden steepening near the
edge, clearly at variance with observed behaviors, e.g.,
Figs.~\ref{fig:3gams} and~\ref{fig:mshgamma}).

\begin{figure}
\centerline{\includegraphics[width=2truein]{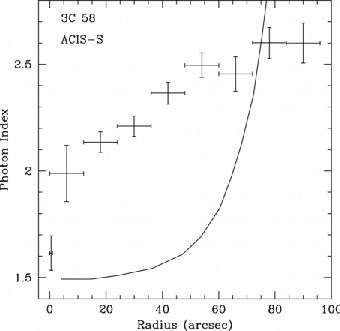}
\hskip0.2truein\includegraphics[width=2truein]{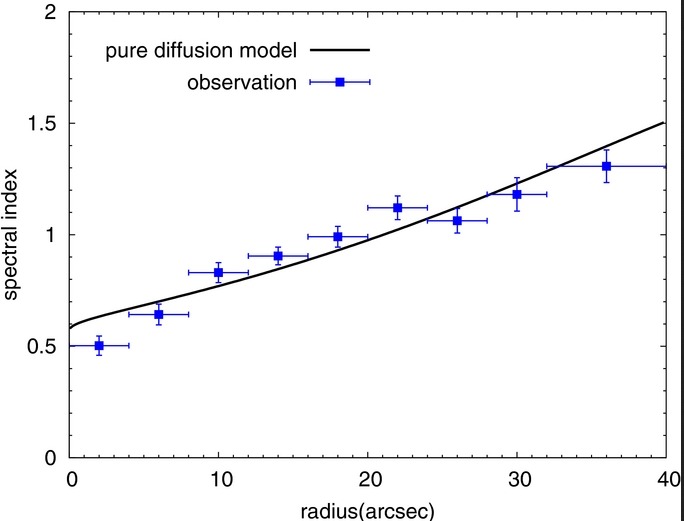}}
\caption{Left: Observed photon index $\Gamma$ as a function of radius in 3C
  58 \citep{slane04}. Curve: Prediction from Kennel-Coroniti pure-MHD
  spherical advection model \citep{reynolds03}. Right:  Diffusion model
for G21.5$-$0.9 \citep[][data from \cite{slane00}]{tang12}.}
\label{fig:diffmodels}
\end{figure}

Simple PWN models offer two possibilities for spectral steepening,
both based on radiative losses of electrons, but invoking different
pictures of particle transport from the wind shock through the nebula.
In pure-advection models like those of \cite{kennel84a} and
\cite{reynolds09}, steepening can be thought of as due to the
shrinkage of effective source size with increasing photon energy, as
energy losses deplete electrons with higher energies at smaller radii.
However, such models all share one important failing: they predict
that the PWN spectral index should be constant throughout the source
until a sudden steepening at the (energy-dependent) edge radius
(Fig.~\ref{fig:diffmodels}).

However, advection models in inhomogeneous sources can produce the
frequently observed steepenings by values $\Delta$ greater than 0.5.
The models can be tested because they achieve such values of $\Delta$
by source shrinkage at related rates; that is, the source size $R$
obeys $R \propto E^m$ with $m$ and $\Delta$ related by the model.  The
Kennel-Coroniti model predicts $\Delta = (4 + \alpha)/9$, where
$\alpha = 0.6$ characterizes the assumed injected spectrum, explaining
optical and higher-energy emission.  A triumph of this model was the
successful explanation ($m = -1/9$) of the rocket observations of the
Crab in the 1970's, as described above.

More extreme source gradients are required to produce $\Delta$
significantly in excess of 0.5, however. \cite{reynolds09} considered
ad-hoc source gradients in flow-tube radius, magnetic field, density,
and velocity, and connected them to predicted values of $\Delta.$ The
large parameter space is significantly constrained if the source
shrinkage with size can actually be measured.  If magnetic-field
strength varies as $r^{m_B}$ and density as $r^{m_\rho}$, conical
flows (jets or spherical winds) with $\alpha = 0$ (for simplicity)
satisfy
\begin{equation}
\frac{\Delta}{-m} = 3 + m_B + m_\rho.
\end{equation}
For G21.5$-$0.9, $\Delta/(-m) = 4.29$ so $m_B + m_\rho = 1.3$,
requiring that either magnetic field $B$ or gas density $\rho$, or
both, {\sl rise} with radius.  This could conceivably happen due to
turbulent amplification of magnetic field and/or mass loading of the
PWN outflow \citep{lyutikov03}, but seems unlikely.  A similar result
holds for \msh, requiring $m_B + m_\rho \sim 1$.  While a value for
$m$ can be determined for the Crab's counter-jet, an independent value
of $\Delta$ for that region alone is difficult to obtain, since that
feature is embedded in the radio nebula and is not morphologically
distinct.

Alternatively, particle transport may occur through diffusion.  Purely
diffusive models \citep{gratton72,reynolds91,tang12} can produce
fairly linear increases in spectal index with radius
(Fig.~\ref{fig:diffmodels}).  However, they do not predict a change of
source size with photon energy, and the spectral breaks must be
intrinsic.  Combination models of diffusion and advection
\citep{tang12} offer more flexibility, but are not amenable to
analytic investigation and have not yet been explored in great detail.
These models all remain quite simple, with spherical or simple
one-dimensional outflow geometries.  The gradual increase of spectral
index with radius probably requires a mix of particles of different
ages at each radius; such a mix can be achieved by diffusion, but also
by more complicated flow geometries, such as the backflows found in
relativistic-MHD simulations such as those of \cite{komissarov04} or
\cite{delzanna04}.

The question of the origin of the full SED of the synchrotron spectrum
from PWNe is thus still unsettled, so the observations cannot yet be
brought to bear directly on theories of particle energization in PWNe.
Even the fundamental origin of the radio-emitting particles is still
mysterious.  Relativistic-MHD simulations still lack the ability to
make detailed spectral predictions, as these require propagating the
full particle spectrum with each fluid element, and are
computationally prohibitive at this time.  It is most likely, however,
that further progress will require the development of this capability.
In the meantime, the observational characterization of PWN SEDs and
spatially resolved spectra remain important tools for constraining
models, and may ultimately contribute essential clues to solutions of
these basic problems.

All three PWNe observed with \nust\ have been detected at TeV
wavelengths.  Gamma-ray observations hold out the promise of further
constraining the particle distributions of PWNe known at other
wavelengths, but also of discovering new types of object that are
predominantly TeV emitters.  Both cases are discussed in 
Section~\ref{sec:gamma}.

%
%

\section{X-ray pulsar wind nebulae outside supernova remnants}
\label{sec:oldpwne}

\subsection{
Expected general properties of PWNe of supersonically moving pulsars}
\label{subsec:general}

\begin{figure}[t]
\sidecaption
\includegraphics[scale=0.6]{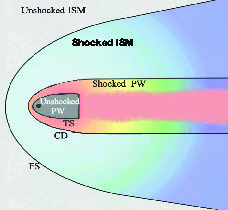}
\caption{Cartoon of a head-tail PWN created by a supersonically moving
  pulsar.  A synchrotron (e.g., X-ray) PWN is produced by the shocked
  PW flowing between the TS and CD surfaces, while the shocked
  circumstellar medium between the FS and CD surfaces is expected to
  be the source of IR-optical-UV radiation.  }
\label{fig:head-tail_sketch}
\end{figure}

For the first 500--1000 years after the SN explosion, the SNR's radius
$R_{\rm snr}$ almost linearly increases with time with a typical speed
of $\sim 10,000$ km s$^{-1}$, much faster than the typical puslar
speed of a few hundred km s$^{-1}$.  At larger ages the SNR expansion
slows down ($R_{\rm snr}\propto t^{2/5}$ and $t^{3/10}$ in the Sedov
and pressure-driven snowplow stages, respectively), while the pulsar
keeps moving with about the same velocity and eventually leaves the
SNR at an age of $\sim 20$--200 kyr (see Fig.\ 4 in
\citealt{arzoumanian02}).
This means that most of the known pulsars are moving in the ISM with
the speed $V_{\rm psr}$ considerably exceeding the speed of sound in
the ambient medium, $c_s=(\gamma_{\rm ad} k T/\mu m_{\rm H})^{1/2}
\sim 3$--30 km s$^{-1}$, where $\gamma_{\rm ad}$ is the adiabatic
index ($\gamma_{\rm ad} = 5/3$ for monoatomic gases), $T$ is the
temperature, and $\mu$ is the molecular weight.  The supersonic motion
of the pulsar drastically changes the PWN morphology (see
\citealt{ref:vds04}, \citealt{gaensler06}, and references therein).
Since the ram pressure $p_{\rm ram}=\rho_{\rm amb} V_{\rm psr}^2 =
1.5\times 10^{-9} n_b (V_{\rm psr}/300\,{\rm km\,s}^{-1})^2$ dyn
cm$^{-2}$ (where $n_b$ is the ambient baryon number density in units
of cm$^{-3}$) exceeds the ambient pressure $p_{\rm amb}$,
\begin{equation}
p_{\rm ram}/p_{\rm amb} = \gamma_{\rm ad}{\cal M}^2\gg 1\,,
\label{ram-to-amb}
\end{equation}
(${\cal M}=V_{\rm psr}/c_s$ is the Mach number), the PWN acquires a
cometary shape with a compact head around the pulsar and a long tail
behind it (Figure~\ref{fig:head-tail_sketch}).  In an idealized picture,
the interaction of the pulsar wind (PW) with the ambient
(circumpulsar) medium creates three distinct regions. The
bullet-shaped cavity around the pulsar is filled with the {\em
  unshocked relativistic PW} confined within the termination shock
(TS).  The {\em shocked PW}, which is the main source of synchrotron
radiation in X-rays, is flowing away between the TS and the contact
discontinuity (CD) that separates the shocked PW from the shocked
ambient medium.  Finally, the compressed and heated {\em shocked
  ambient medium} between the CD and the forward bow shock (FS) is
expected to emit IR-optical-UV radiation in spectral lines and
continuum.

For an isotropic pulsar wind, a characteristic distance from the
pulsar at which the ram pressure of the unshocked PW, $\dot{E}_w/(4\pi
c r^2)$, balances the ram pressure of the ambient medium (the
so-called stagnation point) is
\begin{equation}
R_0=\left(\frac{\dot{E}_w}{4\pi c p_{\rm ram}}\right)^{1/2} = 
1.3\times 10^{16} \dot{E}_{w,35}^{1/2} n_b^{-1/2} (V_{\rm psr}/300\,{\rm km\,s}^{-1})^{-1}\,\, {\rm cm}, 
\label{R0}
\end{equation}
where $\dot{E}_w=\xi_w\dot{E}=10^{35}\dot{E}_{w,35}$ erg s$^{-1}$, $\dot{E}$ is the
pulsar's  spin-down power, and $\xi_w < 1$ is the fraction of $\dot{E}$
that powers the PW.
For an unmagnetized PW, $R_0$ is approximately equal to the distance 
of the CD apex ahead of the pulsar, $R_{\rm cd}\approx R_0$ 
\citep{ref:vds04}.
The characteristic angular separation between the pulsar and the sky projection
of the CD surface is
\begin{equation}
\theta_0 \approx R_0/d = 0\farcs89 d_{1.0}^{-1} \dot{E}_{w,35}^{1/2} n_b^{-1/2} (V_\perp/300\,{\rm km\,s}^{-1})^{-1}\,\sin i,
\label{theta0}
\end{equation}
where $d = 1.0 d_{1.0}$ kpc is the distance\footnote{In this Section
  the subscript of $d$ means the scaling distance in kpc.}, $i$ is
the angle between the line of sight and the pulsar
velocity\footnote{Equation \ref{theta0} is applicable at not too small
  $\sin i$; see \citet{romani2010}.}, and $V_\perp = V_{\rm psr} \sin i$.  The small value of $\theta_0$
implies that subarcsecond resolution (provided only by {\sl Chandra}
among the currently active X-ray observatories) is required to resolve
the PWN head from the pulsar even for nearby pulsars.

The unshocked PW consists of relativistic particles (likely electrons
and positrons) and a magnetic field.  The magnetization parameter
$\sigma$, defined as the ratio of the Poynting flux, $B^2c/(4\pi)$, to
the particle enthalpy flux, is unknown. Pulsar models predict
$\sigma\gg 1$ immediately outside the pulsar magnetosphere, while PWN
models require $\sigma \lesssim 1$ (or even $\sigma\ll1$;
\citealt{kennel84a}) just upstream of the TS.  The decrease of
$\sigma$ with distance from the pulsar could be due to transfer of the
magnetic field energy to the particles, e.g., by magnetic field
reconnection in the striped PW (see \citealt{kirketal09}, and
references therein).  For a given magnetization, the magnetic field
upstream of the TS at the stagnation point can be estimated as
\begin{equation}
B\sim\left[\frac{\dot{E}_w\sigma}{cR_0^2(\sigma+1)}\right]^{1/2} =
\left(\frac{4\pi p_{\rm ram}\sigma}{\sigma+1}\right)^{1/2} \approx
140 \left(\frac{n_b \sigma}{\sigma+1}\right)^{1/2} \frac{V_{\rm psr}}{300\,{\rm km\, s}^{-1}} \,\, \mu{\rm G}\,. 
\label{Bfield}
\end{equation}
The magnetic field can be somewhat higher in the shocked PW, up to a
factor of 3 immediately downstream of TS, at $\sigma \ll 1$
\citep{kennel84a}.  Thus, characteristic PWN magnetic field
values are expected to be of the order of 10--100\,$\mu$G.

Typical energies of synchrotron photons emitted in such fields can be
estimated as
\begin{equation}
E = \zeta\frac{heB_\perp\gamma^2}{2\pi m_e c} = 1.16\zeta B_{-5}\gamma_8^2\,\,{\rm keV} = 4.43 \eta B_{-5} (E_e/100\,{\rm TeV})^2\,\,{\rm keV}\,
\label{synh-energy}
\end{equation}
where $B_\perp = 10^{-5} B_{-5}$ G is the magnetic field component
perpendicular to the electron velocity, $\gamma=E_e/(m_ec^2) = 10^8
\gamma_8$ is the electron Lorentz factor, and $\zeta\sim 1$ is a
numerical factor.  The synchrotron emission spectum is determined by
the electron spectrum, which depends on the still poorly understood
acceleration mechanism.  The commonly considered Fermi acceleration
mechanism at fronts of relativistic shocks (e.g, the TS) gives a
power-law (PL) electron spectrum, $dN_e/d\gamma \propto \gamma^{-p}$
in the range $\gamma_{\rm min} <\gamma <\gamma_{\rm max}$, with
$p\gtrsim 2$ (see, e.g., Chapter 6 of the review by
\citealt{bykovetal2012}).  Such an electron spectrum produces a PL
photon spectrum, $dN/dE \propto E^{-\Gamma}$ in the $E_{\rm min}
<E<E_{\rm max}$ range, with the photon index $\Gamma = (p+1)/2 \gtrsim
1.5$.  The maximum Lorentz factor of accelerated
electrons\footnote{Note that $\gamma_{\rm max}m_ec^2\sim e\Phi
  \sigma/(\sigma+1)$, where $\Phi=B_{\rm LC}R_{\rm LC}\sim B_{\rm
    NS}R_{\rm NS}^3 \Omega^2/c^2$ is the potential drop across the
  pulsar's polar cap, $R_{\rm LC}=c/\Omega$ is the light cylinder
  radius, $\Omega=2\pi/P$, $B_{\rm LC}$ and $B_{\rm NS}$ are the
  magnetic field values at the light cylinder and the neutron star
  surface. The maximum electron energies in other acceleration models
  (e.g., \citealt{romanova05}) are also fractions of $e\Phi$.},
$\gamma_{\rm max} \lesssim (e/m_ec^2)[\dot{E}\sigma/c(\sigma+1)]^{1/2}
\approx 1.1\times 10^9 \dot{E}_{35}^{1/2} [\sigma/(\sigma+1)]^{1/2}$,
can be estimated from the condition $R_g < R_0$, where $R_g=\gamma
m_ec^2/(eB)=1.7\times 10^{16}\gamma_8 B_{-5}^{-1}$ cm is the gyration
radius.  This corresponds to the maximum synchrotron photon energy
\begin{equation}
E_{\rm max} \lesssim 
130 \zeta \dot{E}_{35} B_{-5} \sigma/(\sigma +1)\,\, {\rm keV}.
\label{max-energy}
\end{equation}
This equation shows that one should not expect X-ray PWNe from very old,
low-power pulsars, but head-tail PWNe could be expected at UV-optical-IR
wavelengths.

The qualitative head-tail PWN picture is generally
 confirmed by analytical estimates (e.g., \citealt{romanova05}) and
 numerical simulations. For instance, 
\citet{ref:bucciantini05} presented relativistic MHD 
axisymmetric simulations 
for ${\cal M}=30$ and three values of the PW magnetization $\sigma$
upstream of the TS.  Figure 1 of that paper shows that for an
isotropic PW with a toroidal magnetic field the TS, CD and FS apices
are at distances of $\approx R_0$, $1.3 R_0$ and $1.7R_0$,
respectively, at $\sigma=0.002$, while the radius of the cylindrical
tail (confined by the CD surface behind the pulsar) is $r_{\rm cd}
\approx 4R_0$, almost independent of magnetization.  The bulk flow
velocity in the tail reaches 0.8--$0.9 c$ at its periphery (closer to
the CD), being 0.1--$0.3 c$ in the central channel (behind the TS
bullet).  Simulated maps of synchrotron brightness (Figures 4 and 5 in
\citealt{ref:bucciantini05}) show that the brightness is maximal at
the PWN head, gradually decreasing with distance from the pulsar in
the PWN tail.

The inner and outer channels of the tail flow should mix with each
other at larger distances from the pulsar due to shear instability.
At even larger distances the flow should slow down due to mass-loading
of the ambient matter, which leads to additional tail broadening
(e.g., \citealt{morlino15}).

As the outflowing electrons lose their energy to synchrotron radiation,
we can expect spectral softening with increasing distance from the pulsar,
which means the observed tail's length should increase with decreasing photon
energy (e.g., it should be larger in the radio than in the X-rays).
The length scale of the tail at photon energy $E$ (in keV) can be  estimated as
\begin{equation}
l_{\rm tail} \sim V_{\rm flow}\tau_{\rm syn} \sim 18 (V_{\rm flow}/10,000\,{\rm km\,s}^{-1})\zeta^{1/2} E^{-1/2} B_{-5}^{-3/2}\,\,{\rm pc}\
\label{tail-length}
\end{equation}
where $V_{\rm flow}$ is a typical flow velocity and $\tau_{\rm syn}
=1.6 \gamma_8^{-1} B_{-5}^{-2}\,{\rm kyr} \sim 1.7 \zeta^{1/2} E^{-1/2} B_{-5}^{-3/2}$ kyr is the
synchrotron cooling time. We should note that such an estimate is very crude
because both the flow velocity and the magnetic field vary with the
distance from the pulsar.

The simulations by \citet{ref:bucciantini05} were done for an
isotropic PW.  We, however, know that at least in some young pulsars
the PW is mostly concentrated around the equatorial plane,
perpendicular to the spin axis, and the PWN has equatorial and polar
components (e.g., the torus and jets in the Crab pulsar;
\citealt{weisskopf00}). In this case the ram pressure of the unshocked
PW becomes anisotropic, which changes not only the distance to the
stagnation point but also the overall PWN appearance.  In particular,
the PWN shape strongly depends on the angle between the pulsar's
velocity and spin vectors, as well as on the angle between the spin
vector and the line of sight.  Three-dimensional simulations for
several cases of anisotropic PW were presented by
\citet{ref:vigelius07}, assuming nonrelativistic, unmagnetized flows
(see also \citealt{ref:wilkin00} and \citealt{romani2010} for
analytical approximations). The PW anisotropy should strongly affect
the shape of the PWN head, which may become substantially different
from the bullet-like one.  On the other hand, the shape of tail should
not be so strongly affected, especially at large distances behind the
pulsar.

In addition to synchrotron (and inverse Compton -- see Section 4)
emission from the shocked PW, one can expect emission from the shocked ISM
between the FS and the CD.  
While passing through the FS, it is
compressed and heated 
up to temperatures
\begin{equation}
T \approx (3/16) (\mu m_p/k) V_{\rm psr}^2 =
1.2\times 10^6 (\mu/0.6)(V_{\rm psr}/300\,{\rm km\, s}^{-1})^2\, {\rm K}\, ,
\label{ion-temperature}
\end{equation}
where $\mu m_p$ is the mean 
mass per particle (including electrons; see \citealt{ref:bykov08} for details).
The shocked ISM emits in both
continuum and spectral lines in the optical, UV, and even soft X-ray
ranges. If there are neutral hydrogen atoms in the ambient medium
ahead of the pulsar, they can be excited at the FS and emit spectral
lines in the course of radiative de-excitation.

Thus, a rotation-powered pulsar that has left its parent SNR should be
accompanied by a nebula that consists of a head-tail synchrotron
component, emitting in a broad energy range, perhaps from the radio to
soft $\gamma$-rays, enveloped by a (forward) bow-shock component
observable in the optical-UV (particularly, in the Balmer
lines). Below we will see, however, that observational results do not
always coincide with the predictions of the current simple models.

\subsection{Observational results}
\label{subsec:obs_results}

Although most of the known rotation-powered pulsars have left their
SNRs and are moving with supersonic velocities, their spin-down powers
(and the PWN luminosities) have significantly decreased with age.
Therefore, the number of detected head-tail (bow-shock) PWNe is
relatively small.  Among over 70 PWNe and PWN candidates detected in
X-rays \citep{ref:kargaltsev13}, only about 15 PWNe are certainly
created by old (and/or fast) enough pulsars that have left their
parent SNRs.  These have spin-down powers 
$0.01 \lesssim
\dot{E}_{35}\lesssim 30$
and characteristic (spin-down) pulsar ages
$\tau_{\rm sd}\gtrsim 20$ kyr.  Of the remainder, about 25 -- 30 are
still inside SNRs, and the rest are ambiguous, or not yet confirmed as
PWNe.  An even smaller number of head-tail PWNe were observed with
exposures deep enough to accurately measure the X-ray PWN
properties. Below we present a few examples of such observations.

\subsubsection{The Mouse: A textbook example? 
\label{subsubsec:mouse}}

The Mouse PWN was discovered in a radio survey of the Galactic center
region \citep{yusefzadeh1987}. The VLA image showed a bright compact
head (``snout''), a bulbous $\sim2'$ long ``body'', and a remarkably
long, $\sim 12'$, narrow ``tail'' (see the right panels in Figure
\ref{fig:mouse-radio}). \citet{camiloetal2002} discovered a 98 ms pulsar,
J1747--2958, within the Mouse's head, with a spin-down power
$\dot{E}=2.5\times 10^{36}$ erg s$^{-1}$ and a characteristic age
$\tau_{\rm sd}=25$ kyr.  \citet{halesetal2009} measured the proper
motion of the radio PWN head\footnote{It can differ from the pulsar's
  proper motion if the pulsar moves in a non-uniform ambient medium,
  but we will neglect this difference here.}, $\mu=12.9\pm 1.8$ mas
yr$^{-1}$, which corresponds to a transverse velocity $V_\perp =
(306\pm 43)d_5$ km s$^{-1}$, where $d_5=d/5\,{\rm kpc}$.  Based on the
projected tail length, $\sim 17d_5$ pc, and the lack of an SNR that
could possibly be associated with the pulsar (see, however,
\citealt{yusefzadeh-gaensler2005}), \citet{halesetal2009} argue that
the true age of the pulsar is $\gtrsim 160$ kyr.

\begin{figure}[t]
\sidecaption
\includegraphics[scale=0.22]{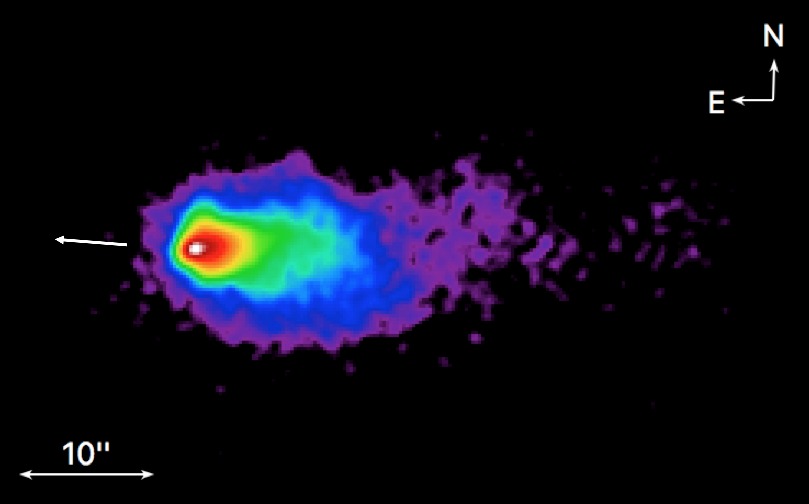}
\includegraphics[scale=0.3]{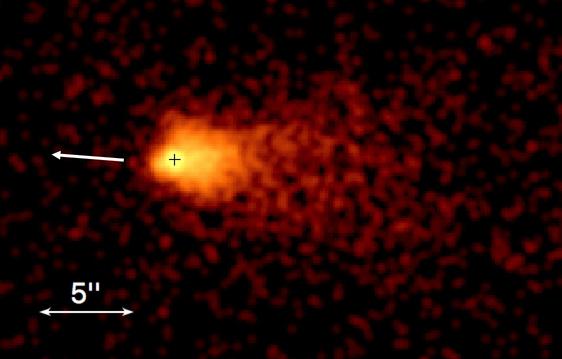}
\caption{Left: Merged image from 5 {\sl Chandra} ACIS observations of
  the Mouse (0.5--8 keV, 154 ks exposure).  Right: {\sl Chandra} HRC
  image of the Mouse (58 ks exposure).  The cross in the right image
  marks the pulsar position; the arrows show the direction of proper
  motion.  }
\label{fig:mouse-acis}
\end{figure}

\begin{figure}
\includegraphics[scale=1.15]{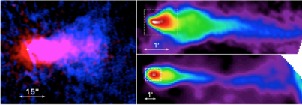}
\caption{Left: Composite X-ray (red; ACIS) and radio (blue; VLA, 1\farcs07 beam)
image of the Mouse PWN. The pulsar position is shown by the white cross.
Right: VLA radio images (top: $11''$ beam, bottom: $32''$ beam) showing the extended tail of the Mouse. The field of view of the left image is shown by the
dashed white box in the lower right panel.
The radio images were obtained from the NRAO VLA Archive.}
\label{fig:mouse-radio}
\end{figure}

\begin{figure}
\sidecaption
\includegraphics[scale=0.3]{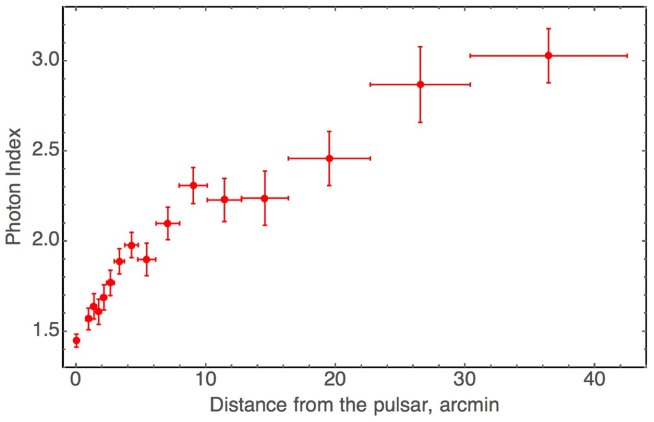}
\caption{Photon index $\Gamma$ as a function of distance from the pulsar
along the tail of the Mouse (Klingler et al.\ 2017, in preparation).}
\label{fig:mouse-spectrum}
\end{figure}

The Mouse PWN was observed with {\sl Chandra} by
\citet{gaenslermouse2004} and Klingler et al.\ (2017, in preparation), with
36 ks and 120 ks exposures, respectively.  These observations have
shown an X-ray nebula with a compact bright head and a tail of $\sim
45''$ length, a factor of 16 shorter than the radio tail (see Figure
\ref{fig:mouse-acis}). The X-ray luminosity of the PWN, $L_{\rm
  0.5-8\,keV}\approx 3.3\times 10^{34}d_5^2$ erg s$^{-1}$, is about
$0.013 d_5^2$ of the pulsar's spin-down power, unusually high compared
to other PWNe \citep{kargaltsevpavlov2008}.  The distance from the
pulsar to the projected leading edge of the head, $\theta_0\approx
1''$, corresponds to $n_b\sim 0.8 \xi_w d_5^{-4} \sin^2i$ cm$^{-3}$
(see Equation \ref{theta0}), typical for a warm phase of the ISM
(e.g., \citealt{ref:bykov08}). For
a sound speed of $\sim 10$ km s$^{-1}$, expected for this phase, the
pulsar's Mach number can be estimated as ${\cal M}\sim 30 d_5/\sin i$.
The spatially-resolved X-ray spectrum showed a significant increase of
the photon index with increasing distance from the pulsar, from
$\Gamma\approx 1.6$ in the immediate vicinity of the pulsar to
$\Gamma\approx 3.0$ at $\sim 40''$ from the pulsar (see Figure
\ref{fig:mouse-spectrum}).  The spectral softening could be caused by
synchrotron cooling.  Assuming equipartition between the magnetic field
energy and the kinetic energy of relativistic electrons, the magnetic 
field strength estimated from the observed X-ray emission is a few 
hundred $\mu$G, substantially higher than in other head-tail PWNe\footnote{We should note, however, that such estimates assume that the magnetic field
strength is about the same throughout the emitting volume, which may be far from reality.}.
For such a magnetic field the projected length of the X-ray tail, 
$l_{\rm tail} \sim 1\, d_5$ kpc, corresponds to a flow velocity
$V_{\rm flow} \sim 20,000\, B_{-4}^{3/2}\,d_5/\sin i$ km s$^{-1}$, much
higher than $V_{\rm psr}$ but significantly lower than 
the mildly relativistic speeds predicted by
\citet{ref:bucciantini05}. Moreover, the comparison of the X-ray and
high-resolution radio images (see the left panel of Figure
\ref{fig:mouse-radio}) suggests that the flow is faster in the middle of
the tail, contrary to the model predictions.

The radio tail of the Mouse is longer than in any other known PWN.
Just behind the pulsar, the radio image looks like a cone with an
$\approx 25^\circ$ half-opening angle (much broader than the Mach cone
at the above-estimated ${\cal M}\sim 30$) and a vertex at the pulsar
position.  The cone abruptly narrows at $\sim 1'$ from the pulsar.
Such behavior is not explained by the current PWN models.  The Mouse
is one of the few PWNe with a mapped radio polarization.  Polarization
mesurements by \citet{yusefzadeh-gaensler2005} suggest that the
magnetic field wraps around the bow shock structure near the apex of
the system, but runs parallel to the direction of the pulsar's motion
in the tail behind the pulsar. Such a magnetic field distribution is
different from the toroidal one assumed in the models by
\citet{ref:bucciantini05}.  Thus, the Mouse has a few features
consistent with the model predictions, but the models do not fully
agree with the observations, particularly in the radio.

\subsubsection{Geminga: An odd ``three-tail'' PWN}
\label{subsubsec:geminga}

\begin{figure}[t]
\sidecaption
\includegraphics[scale=0.45]{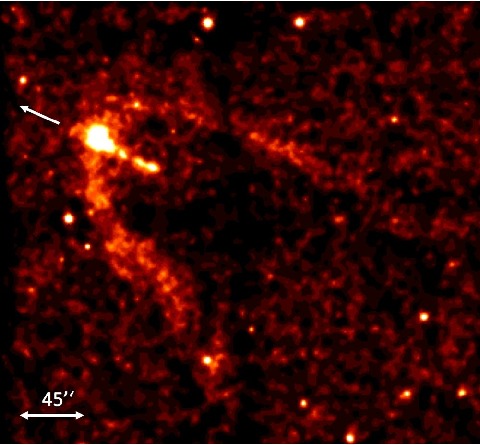}
\caption{Combined {\sl Chandra} ACIS image of the Geminga PWN (0.5-8 keV, 540 ks). The arrow shows
the direction of the pulsar's proper motion. }
\label{fig:geminga}
\end{figure}

\begin{figure}
\includegraphics[scale=0.7]{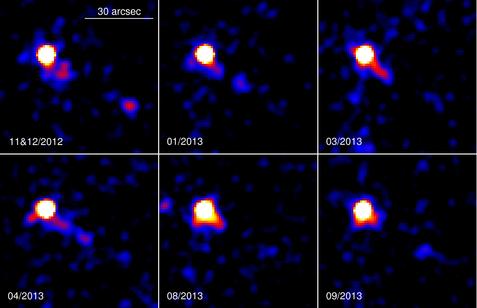}
\caption{Images of the central tail of the Geminga PWN in separate 
{\sl Chandra} observations, in the 0.5--8 keV band.}
\label{fig:central-tail}
\end{figure}

The X-ray PWN created by the radio-quiet $\gamma$-ray pulsar Geminga
($P=237$ ms, $\dot{E}=3.3\times 10^{34}$ erg s$^{-1}$, $\tau_{\rm
  sd}=340$ kyr, $d=0.25^{+0.23}_{-0.08}$ kpc) looks quite different
from the Mouse\footnote{One should bear in mind, however, that much
  smaller spatial scales can be probed in the nearby Geminga PWN than
  in the Mouse.}  and from the predictions of PWN models.  The proper
motion of this pulsar, $\mu=178.2\pm 1.8$ mas yr$^{-1}$
\citep{faherty2007}, corresponds to the transverse velocity $V_\perp =
(211\pm 2) d_{0.25}$ km s$^{-1}$.  Observations with {\sl XMM-Newton}
revealed two bent ``tails'' behind the pulsar, on both sides of its
sky trajectory \citep{caraveo2003}, while {\sl Chandra} observations
with higher spatial resolution showed a shorter third tail between
the two lateral tails (\citealt{pavlov2006}; \citealt{pavlov2010}).
The most detailed data on the Geminga PWN were provided by a series of
12 {\sl Chandra} observations carried out in 2012--2013, with a total
exposure of about 580 ks \citep{posselt2016}.  Figure \ref{fig:geminga}
shows a summed image from these observations, where we see two lateral
tails of $\sim 3'$ ($0.2 d_{0.25}$ pc) length and one $\sim0.45''$
($0.05 d_{0.25}$ pc) long central tail. Surprisingly, there is only a
hint of bow-like emission ahead of the pulsar and no bright, filled
`bullet' predicted by the PWN models assuming an isotropic PW. The
0.3--8 keV luminosities of the northern and southern lateral tails,
and the central tail are 1.6, 2.6, and 0.9 $\times 10^{29}d_{0.25}^2$
erg s$^{-1}$, respectively, i.e., the total PWN luminosity is a
fraction of $1.5\times 10^{-5}d_{0.25}^2$ of the pulsar's spin-down
power, three orders of magnitude smaller than for the Mouse. Images from
separate exposures show that the central tail is formed by isolated
``blobs'' seen at different distances from the pulsar in different
observations (see Figure \ref{fig:central-tail}). However, there is no
evidence of constant or decelerated motion of the blobs. The spectra
of the lateral tails are very hard, $\Gamma\approx 0.7$--1.0, much
harder than the spectrum of the central tail, $\Gamma\approx 1.6$, and
they do not show significant changes with increasing distance from the
pulsar.

The nature of the three tails is not certain yet. One could assume
that the lateral tails represent a limb-brightened paraboloid shell of
shocked PW downstream of the TS and their unusually hard spectrum is
emitted by electrons accelerated by the Fermi mechanism at the shocks
that form in two colliding flows (in the reference frame of CD) -- the
PW and the oncoming ambient medium.  However, a lack of diffuse
emission in between the lateral tails strongly suggests that the shell
emissivity is not azimuthally symmetric with respect to the shell axis
(i.e., the direction of motion). Such an asymmetry could be caused by
a strong asymmetry of the PW (e.g., because the equatorial plane
around the pulsar spin axis, where the PW is presumbly concentrated, is
strongly misaligned with the direction of motion) or an azumuthally
asymmetric magnetic field in the shell (see \citealt{posselt2016} for
details).  Alternatively, the lateral tails could be interpreted as
strongly collimated polar outflows (jets) bent by the ram pressure of
the oncoming ambient medium.  The blobs in the central tail could be
short-lived plasmoids formed by magnetic field reconnection in the
relativistic plasma behind the moving pulsar, resembling the
magnetotails of the Solar system planets. Observations at different
wavelengths could clarify the PWN nature, but the Geminga PWN was not
detected in the radio or H$\alpha$.  For any interpretation, we can
conclude that the Geminga's PW is strongly anisotropic, and new models
are required to explain the morphologies and spectra of such PWNe.

\subsubsection{The Guitar: First example of a misaligned outflow}
\label{subsubsec:guitar}

\begin{figure}[t]
\includegraphics[scale=1.0]{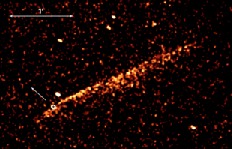}
\caption{Combined {\sl Chandra} ACIS image
(0.5--8 keV, 195 ks total exposure) of PSR B2224+65 and its misaligned outflow,
in a coordinate frame moving with the pulsar.  The image is a combination of
two images separated by an interval of 6 years, during which the pulsar 
moved $1''$ on the sky.
}
\label{fig:guitar-xray}
\end{figure}

\begin{figure}[]
\centering
\includegraphics[scale=0.6]{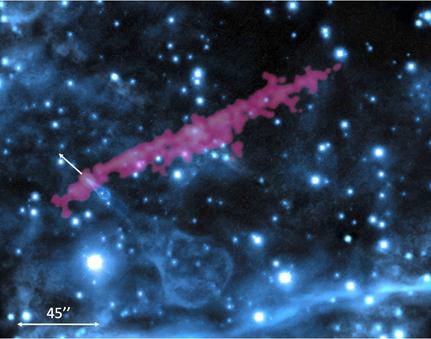}
\caption{Smoothed X-ray image from Figure \ref{fig:guitar-xray} overlaid onto
the H$\alpha$ image
showing the Guitar. The H$\alpha$ image is taken from
 http://chandra.harvard.edu/photo/2015/archives/.
}
\label{fig:guitar-Halpha-xray}
\end{figure}

The Guitar nebula is produced by the relatively old, low-power pulsar
B2224+65 ($P=683$ ms, $\dot{E}=1.2\times 10^{33}$ erg s$^{-1}$,
$\tau_{\rm sd}=1.12$ Myr). The pulsar is among the highest velocity
neutron stars known; its proper motion, $\mu=182\pm3$ mas yr$^{-1}$,
corresponds to the transverse velocity $V_\perp = 860$--1730 km
s$^{-1}$ (the uncertainty is caused by the uncertain distance,
$d=1$--2 kpc).  The guitar-shaped H$\alpha$ nebula was discovered by
\citet{cordes1993} and further studied in several papers (see
\citealt{dolchetal2016} and references therein).  One could expect a
head-tail X-ray PWN within the H$\alpha$ bow shock, but a
high-resolution observation with {\sl Chandra} showed instead a
straight $2'$ ($0.3 d_1$ pc) long feature inclined by $118^\circ$ to
the direction of the pulsar's proper motion (\citealt{wongetal2003};
\citealt{hui-becker2007}; see Figures \ref{fig:guitar-xray} and \ref{fig:guitar-Halpha-xray}).  A second
        {\sl Chandra} observation 6 years later showed that the sharp
        leading edge of the jet-like feature had the same proper
        motion as the pulsar, and it provided evidence for the
        presence of a counter-feature, albeit substantially shorter
        and fainter than the main one \citep{johnson-wang2010}. The
        feature shows a power-law spectrum with $\Gamma\approx 1.6$,
        comparable to that of the point-like source (the pulsar plus
        an unresolved PWN?).  The luminosity of the feature, $L_{\rm
          0.3-7\,keV} \sim 7\times 10^{30} d_1^2$ erg s$^{-1}$,
        exceeds that of the point-like source by a factor of 3--4, and
        is a fraction of $\sim 6\times 10^{-3}d_1^2$ of the pulsar's
        spin-down power.

The lack of a resolved X-ray head-tail PWN could be explained by the very high 
pulsar velocity and low spin-down power. Indeed, according to Equation \ref{R0}, the characteristic PWN size ahead of the pulsar, $R_0\sim 5\times 10^{14} \xi_w^{1/2} n_b^{-1/2} d_1^{-1}$ cm, correponds to the  angular distance\footnote{\citet{chatterjee-cordes2002} estimated $\theta_0=0\farcs06\pm0\farcs02$ from
modeling an H$\alpha$ image obtained with the {\sl Hubble Space Telecope}.}
 as small
as $\theta_0 \sim 0\farcs03 (\xi_w/n_b)^{1/2} d_1^{-2}$, much smaller than the angular
resolution of {\sl Chandra}. The lack (shortness) of the X-ray tail could 
be due to a high magnetic field (hence fast synchrotron cooling)
 in the shocked PW associated with the small
stand-off distance (see Equation \ref{Bfield}). Another reason could be
a low maximum energy of accelerated electrons at the low spin-down power
of B2224+65 (see Equation \ref{max-energy}).

The nature of the elongated feature remains unclear. It might be a pulsar jet,
but such a jet should be
bent by the ram pressure of the
 oncoming ambient
medium while no bending is observed.
\citet{bandiera2008} suggested that the feature is produced by synchrotron 
radiation of highest energy electrons ($\gamma\sim 10^8$) accelerated at the
TS and leaked into the 
ISM along its magnetic field. This scenario,
however, requires a rather high ambient magnetic field ($\sim 45\,\mu{\rm G}$,
according to \citealt{bandiera2008})
 and it remains
unclear why the counter-feature is so much fainter than the main one.
Since the true nature of the feature is not certain yet, we will call it
simply a {\em misaligned outflow}. 
It should be emphasized that, most likely, the misaligned outflow is not 
a (magneto)hydrodynamical flow but rather a stream of high-energy particles
not interacting with each other and with the ISM gas.

\subsubsection{J1509--5850: Another misaligned outflow, a ``three-tail''
compact nebula, and a long tail }
\label{subsubsec:j1509}

\begin{figure}
\includegraphics[scale=0.31]{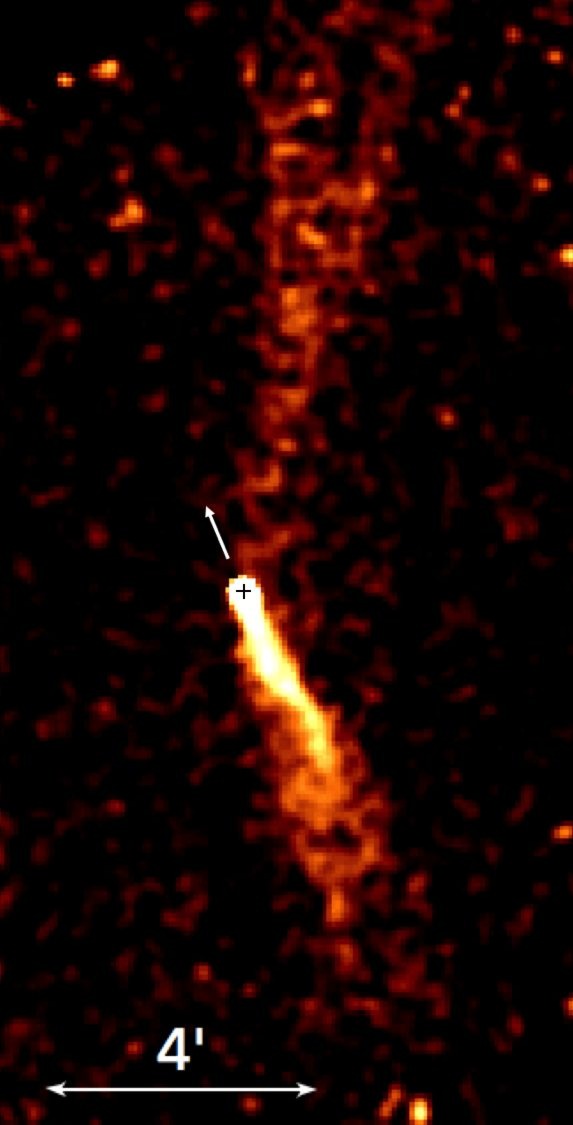}
\includegraphics[scale=0.31]{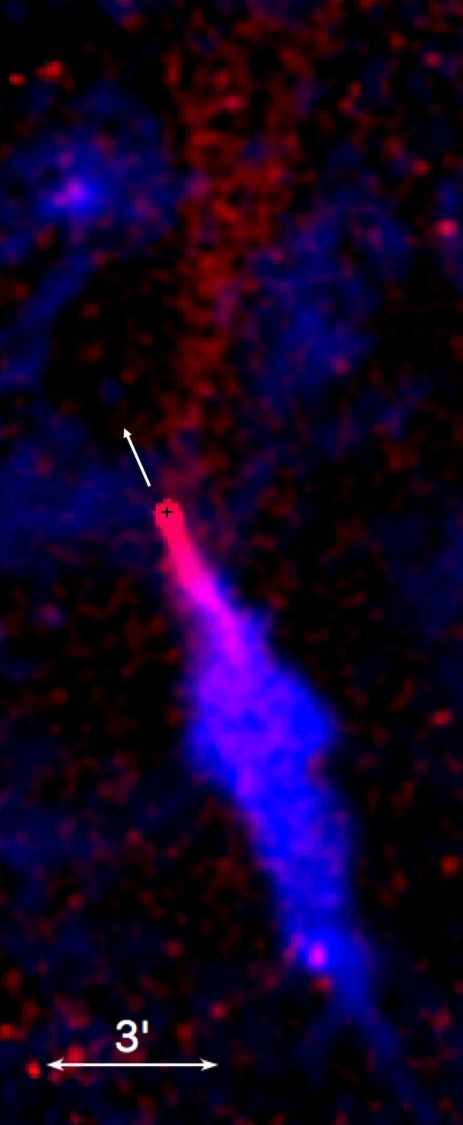}
\caption{Left: {\sl Chandra} ACIS image of the J1509--5850 PWN (0.5--8 keV, 374 ks) showing the southwest tail and the misaligned outflow toward the north.
Right: Combined {\sl Chandra} ACIS (red) and VLA (blue) image of the same PWN.
The arrows show an assumed direction of proper motion.}
\label{fig:J1509-large}
\end{figure}

\begin{figure}
\includegraphics[scale=0.22]{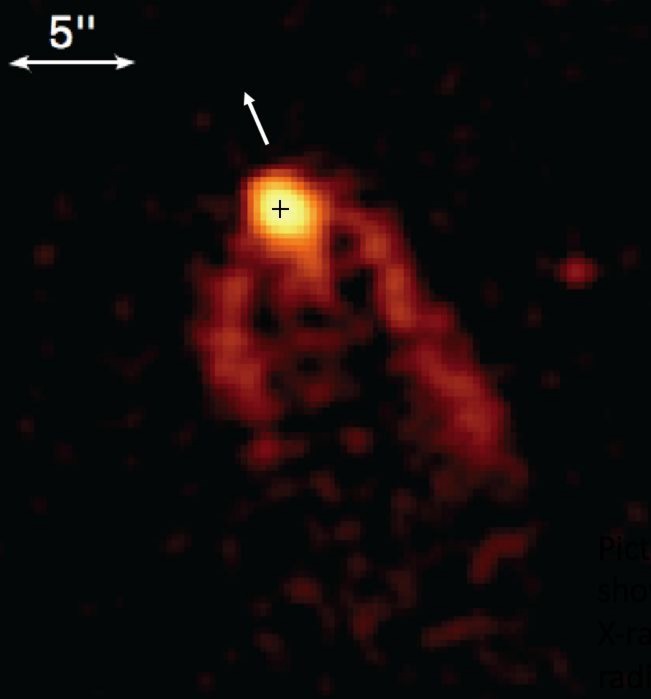}
\includegraphics[scale=0.22]{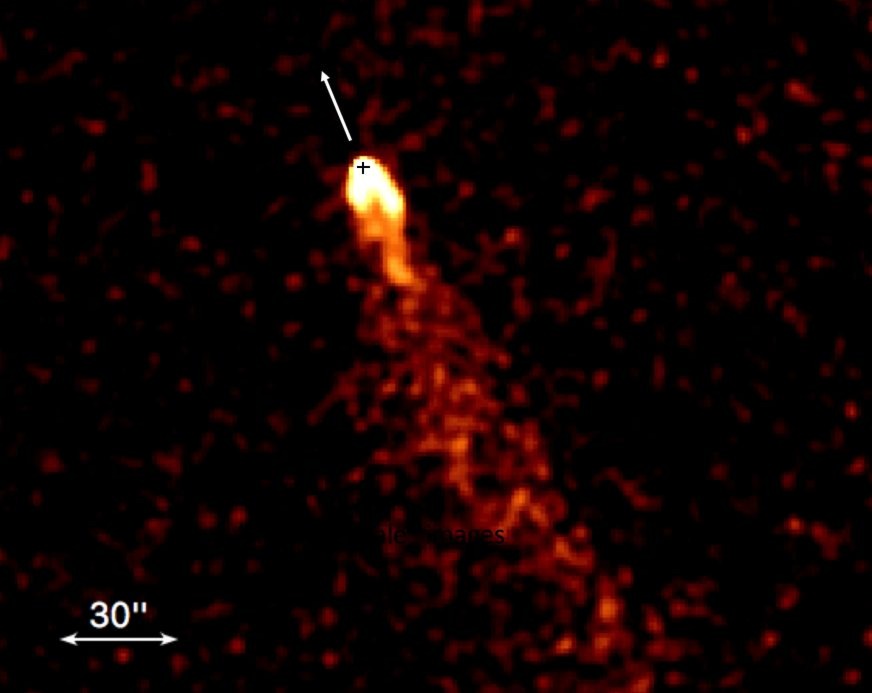}
\caption{Left: Compact X-ray nebula in the vicinty of PSR J1509--5850. Right: 
ACIS image demonstrating the transition from the compact nebula to the southwest tail. }
\label{fig:J1509-CN}
\end{figure}

PSR J1509--5850 is a middle-aged ($\tau_{\rm sd}=154$ kyr) pulsar with 
$P=89$ ms, $\dot{E}=5.1\times 10^{35}$ erg s$^{-1}$, and a dispersion-measure
distance $d\approx 3.8$ kpc. 
Its X-ray PWN, consisting of a compact ``head'' and a 
long
``tail'' 
southwest of the pulsar, was discovered in a {\sl Chandra} observation by \citet{kargaltsev2008}. 
Deep follow-up {\sl Chandra} observations (374 ks total exposure) are
described by \citet{klinglerJ1509-2016}. In addition to the previously detected
southwest tail extending up to $7'$ ($7.7 d_{3.8}$ pc), these observations revealed 
similarly long (but fainter)
 diffuse emission stretched toward the north and the fine 
structure of the PWN ``head'' (see Figure \ref{fig:J1509-large}).
The ``head'' (dubbed the Compact Nebula [CN]
by \citealt{klinglerJ1509-2016}) is resolved into two lateral tails
and one short central tail (Figure \ref{fig:J1509-CN}), remarkably similar to the Geminga PWN.
 Although the pulsar's proper motion
has not been measured, the overall CN and southwest tail morphology 
provides strong evidence
that the pulsar is moving northeast. In this case the northern structure
is another example of a misaligned outflow.

\citet{klinglerJ1509-2016} 
estimated upper and lower limits for the transverse velocity,
$V_\perp \lesssim 640 d_{3.8}$ km s$^{-1}$ and
$V_\perp \gtrsim 160 n_b^{-1/2} d_{3.8}^{-1}$ km s$^{-1}$,
using upper limits on the pulsar's proper motion and stand-off distance
(the latter estimate assumes an isotropic PW). 
Being morphologically similar to the Geminga PWN, the CN of J1509--5850
is a factor of $\sim 200$
 more luminous (e.g., $L_{\rm 0.5-8\, keV}\approx 7.5\times 10^{31}d_{3.8}^2$
erg s$^{-1}$ for the CN lateral tails)
and a factor of
$\sim 10$ more X-ray efficient ($\eta_{\rm 0.5-8\, keV}\equiv L_{\rm 0.5-8\, keV}/\dot{E} \approx 1.5\times
10^{-4}d_{3.8}^2$ vs.\ $1.5\times 10^{-5}d_{0.25}^2$ for Geminga). In addition, the spectra of
the lateral tails are much softer in the CN than in the Geminga PWN 
($\Gamma\approx 1.8$ vs.\ $\Gamma\approx 1$, respectively). The reason
of these differences is currently unclear. 

Being aligned with the CN symmetry axis,
the extended 
tail southwest of the pulsar
 is obviously composed of a shocked PW collimated by 
ram pressure. Its luminosity is $L_{\rm 0.5-8\,keV}\approx 1\times 10^{33}d_{3.8}^2$
erg s$^{-1}$, and its spectral slope, $\Gamma\approx 1.9$,
 does not show any increase (rather a hint of decrease) with increasing
distance from the pulsar. The lack of spectral softening suggests a 
very high speed of the outflowing matter.
Alternatively,
there could be some ``reheating'' 
due to in situ conversion of
magnetic
field
energy into particle energy, e.g., via turbulent
processes and accompanying reconnection, which might
explain the hint of spectral hardening at large distances from
the pulsar. 
The 
tail is also seen in  radio up to about $10'$ from the pulsar \citep{ngJ1509-2010}.
Surprisingly, the radio emission
brightens with distance from the pulsar (contrary to the Mouse tail),
 becomes broader
than the X-ray emission beyond $\approx 3'$,
and then narrows again beyond $\approx 5'$ (see Figure \ref{fig:J1509-large}).
Another difference from the Mouse tail is the predominant magnetic
field orientation, stretched along the tail in the Mouse and 
helical, with the helix axis parallel to the pulsar's direction of
motion,  in
the J1509 
tail. The different magnetic field geometries
possibly reflect different spin-velocity alignments of the parent
pulsars.

The median of the
$7'$ long wedge-like northern structure is inclined
to the CN symmetry axis (alleged direction of proper motion)
by $\approx33^\circ$. Its luminosity, $L_{\rm 0.5-8\, keV}\approx 4\times 10^{32}d_{3.8}^2$, is a factor of 2.5 lower than that of the southwest tail, while
the spectral slope is about the same, with a slight hint of softening 
with increasing distance from the pulsar. 

\subsubsection{The 
complex PWN created by PSR B0355+54}
\label{subsubsec:b0355}

\begin{figure}
\includegraphics[scale=0.22]{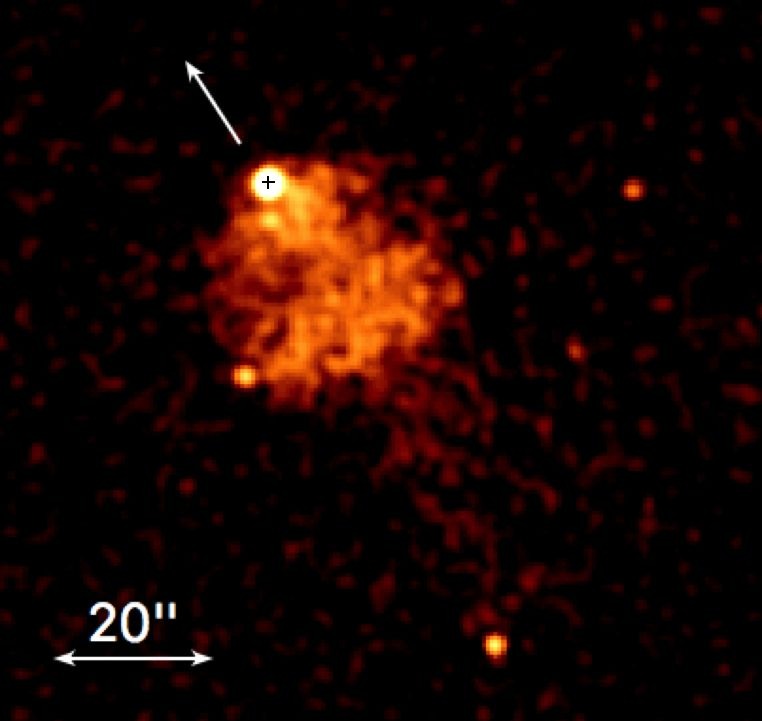}
\includegraphics[scale=0.25]{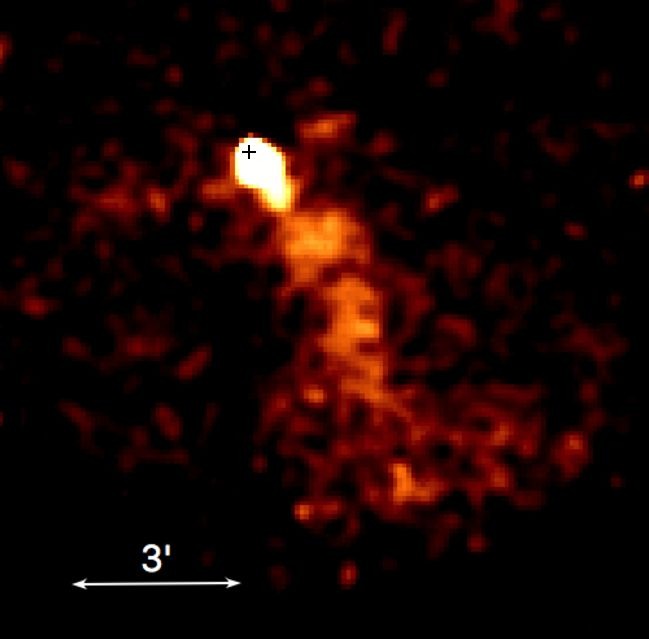}
\caption{{\sl Chandra} ACIS images of the B0355+54 PWN (0.5--8 keV, 395 ks).
Left: The compact PWN in the pulsar vicinity (the Mushroom). Right: The large-scale PWN, including the diffuse tail and the ``whiskers''. The arrows show the
direcion of the proper motion. }
\label{fig:B0355}
\end{figure}

PSR B0355+54, located at a parallax distance of $d=1.0\pm 0.2$ kpc,
is a middle-aged radio pulsar ($\tau_{\rm sd} = 560$ kyr) with a spin-down
power $\dot{E}=4.5\times 10^{34}$ erg s$^{-1}$ and a period $P=156$ ms.
Its 
transverse velocity, $V_\perp = 61^{+12}_{-9} d_1$ km s$^{-1}$ towards
the northeast, is among the
lowest observed.
Observations with {\sl Chandra} and {\sl XMM-Newton} revealed
the presence of a PWN (dubbed the Mushroom by \citealt{kargaltsevpavlov2008})
 consisting of a compact 
``cap'' and a dimmer 
``stem'', with a hint of extended emission visible up to $\sim 7'$ (2 pc) 
southwest of the pulsar
\citep{mcgowan2006}.
A series of 8 {\sl Chandra} observations, performed over an 8-month
period in 2012-2013
(total exposure of 395 ks)
revealed the detailed structure of the B0355+54 PWN
(see Figure \ref{fig:B0355}) and allowed us to measure
the spectra of its elements \citep{klinglerB0355-2016}. In particular,
they showed a ``filled'' morphology of the cap, in contrast with
the ``hollow'' morphologies of the Geminga
PWN and the CN of the J1509--5850 PWN. The cap has a sharp trailing edge
behind the pulsar and is brightened along the axis; its spectral slope
is $\Gamma\approx 1.5$, a typical value for a PWN head. 
The stem is split
into two structures that apparently
originate from the pulsar and slightly diverge
from each other further away. \citet{klinglerB0355-2016} speculate that
these structures could be pulsar's jets swept back by the ram pressure,
which could also explain the brightening along the cap's axis.
Overall, the ``mushroom'' morphology suggests a small angle between
the pulsar's spin axis and our line of sight, in agreement with the lack
of $\gamma$-ray pulsations. 
The cap and stem luminosities are $1.8\times 10^{31}$ and $2.6\times
10^{30}$ erg s$^{-1}$, respectively. 

A long diffuse tail behind the ``mushroom'' is likely due to
synchrotron emission of the shocked PW behind the pulsar.  Its
luminosity is about $3.8\times 10^{31}$ erg s$^{-1}$, more than a half
of the total PWN luminosity, $L_{\rm 0.5-8\,keV} \approx 6.4\times
10^{31}$ erg s$^{-1}$ that corresponds to a total PWN efficiency
$\eta_{\rm 0.5-8\,keV}\approx 1.4\times 10^{-3}$.  The spectrum of the
tail, with a slope $\Gamma\sim 1.7$--1.8, shows only a slight hint of
cooling with increasing distance from the pulsar. This implies either
a fast flow speed (or a very low magnetic field), or particle
re-acceleration within the tail.

The deep observation also allowed \citet{klinglerB0355-2016} to detect
two additional very faint, extended features (dubbed ``whiskers'') on
either side of the compact nebula, likely another example of a
misaligned outflow.

Thus, the B0355+54 PWN shows a particularly rich structure, which remains to be explained by PWN models. 
Radio and H$\alpha$ observations could shed light on its nature, but the PWN
has not been detected at these wavelengths.

\subsubsection{J1741--2054: Another tail behind a nearby middle-aged
pulsar}
\label{subsubsec:j1741}

\begin{figure}
\includegraphics[scale=0.183]{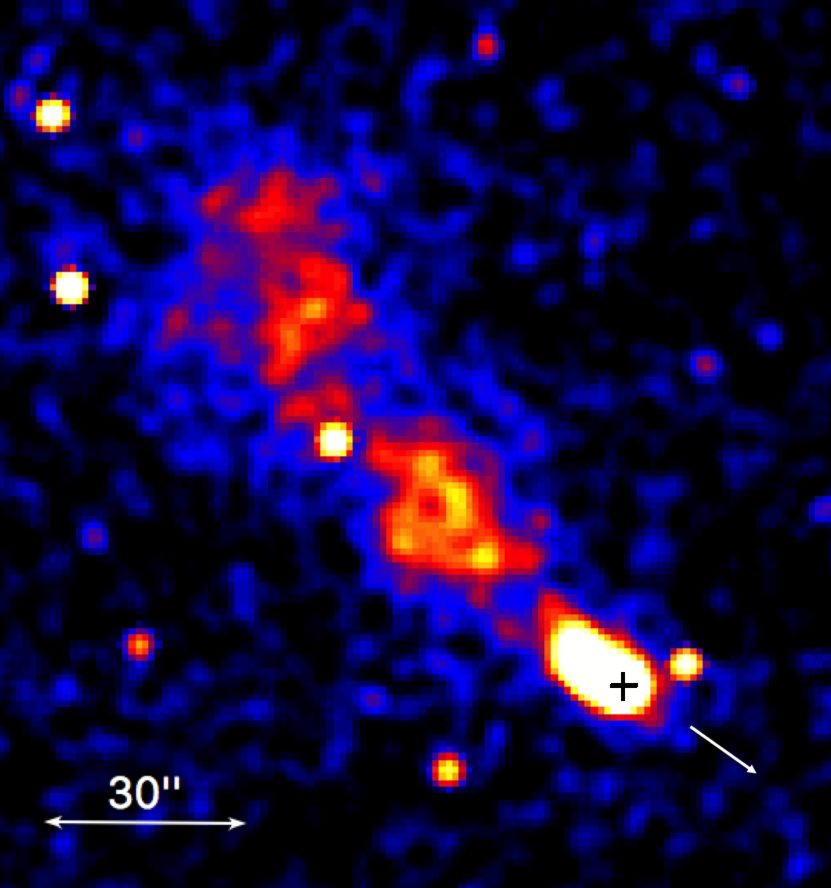}
\includegraphics[scale=0.29]{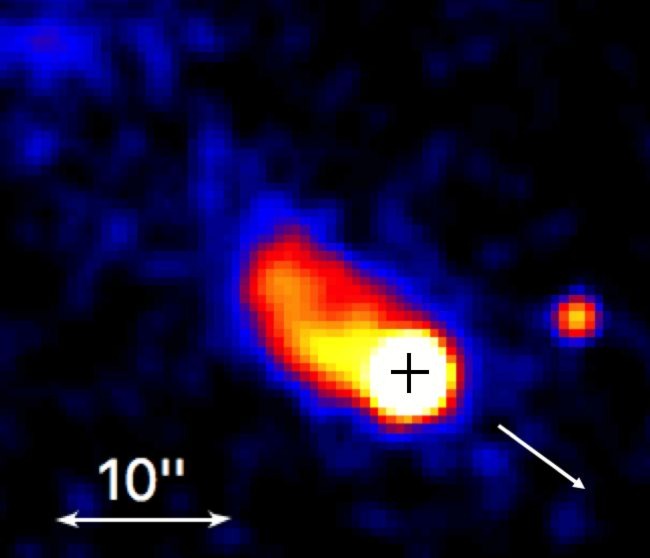}
\caption{Summed 0.3--6 keV {\sl Chandra} ACIS images of the entire 
J1741--2054 PWN
(left) and the leading bright component (right) from 5 observations
in 2013 (282 ks total exposure). Black crosses show the pulsar position,
white arrows show the direction of proper motion. 
}
\label{fig:J1741}
\end{figure}

Another example of a tail-like structure behind a moving pulsar is shown
in Figure \ref{fig:J1741}. This PWN was discovered by \citet{romani2010} and
investigated in detail by \citet{auchettl2015} using results from 6 
{\sl Chandra} observations carried out in 2013 (282 ks total exposure). It is created by the
nearby ($d\sim 0.38$ kpc) middle-aged ($\tau_{\rm sd}=390$ kyr) pulsar
J1741--2044 ($P=413$ ms, $\dot{E}=9.5\times 10^{33}$).
The pulsar's proper motion $\mu=109\pm 10$ mas yr$^{-1}$, measured by
\citet{auchettl2015} from the X-ray images, corresponds to the transverse
velocity $V_\perp = (196\pm 18) d_{0.38}$ km s$^{-1}$. In Figure \ref{fig:J1741} we can
see a tail-like structure in the direction opposite to that of the
proper motion. The structure consists of a compact elongated nebula
of $\sim 15''$ length
and a fainter diffuse tail seen up to $\sim 1\farcm7$ ($\sim 0.2d_{0.38}$ pc)
 from the
pulsar. The tail is slightly bent and apparently consists of two
``lobes''. No small-scale structure (PWN head) is resolved around the
pulsar. The 0.5--10 keV luminosities of the compact nebula and the 
entire PWN, about $5\times 10^{29}$ and $3\times 10^{30}$ erg s$^{-1}$
at $d=0.38$ kpc, are $\sim 5\times 10^{-5}$ and $\sim 3\times 10^{-4}$
of the pulsar's spin-down power. The spectra of the PWN elements are
described by a PL model with $\Gamma \approx 1.5$--1.7, with only a hint
of spectral softening with increasing distance from the pulsar.  

\begin{figure}[h!]
\sidecaption
\includegraphics[scale=0.43]{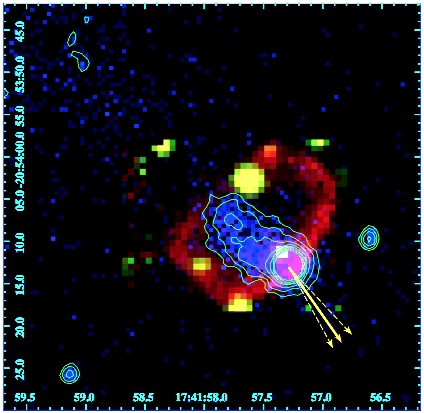}
\caption{The H$\alpha$ nebula 
(red) and the leading tail
 component of the X-ray PWN created by PSR J1741--2051 \citep{auchettl2015}}.
\label{fig:J1741-Halpha}
\end{figure}

For an isotropic PW, one could expect a PWN head with a leading edge
at an angular distance $\theta_0\approx 1'' d_{0.38}^{-2} n_b^{-1/2}
\sin i$ ahead of the pulsar, too small to resolve separately from the
pulsar's PSF.  The actual distance to the stagnation point is likely
even smaller, as seen from the H$\alpha$ image (see Figure
\ref{fig:J1741-Halpha}).  The flat front of the H$\alpha$ bow shock
allows one to assume that the wind of this pulsar is originally
concentrated in the plane perpendicular to the pulsar's velocity,
presumably the equatorial plane, which implies that the pulsar's
rotational axis is parallel to the velocity vector.  Being deflected
by the ram pressure, the shocked PW forms the tail seen as a brighter
compact component (Figure \ref{fig:J1741}, right).  The matter flowing
in the tail is likely decelerated by the ISM entrainment, which leads
to the broadening seen in the diffuse longer tail.  To confirm this
interpretation, flow velocities in the compact and extended tail
components should be measured, but it was not possible with the data
available.

\subsubsection{J0357+3205: A tail detached from the pulsar}
\label{subsubsec:j0357}

\begin{figure}
\includegraphics[scale=0.2]{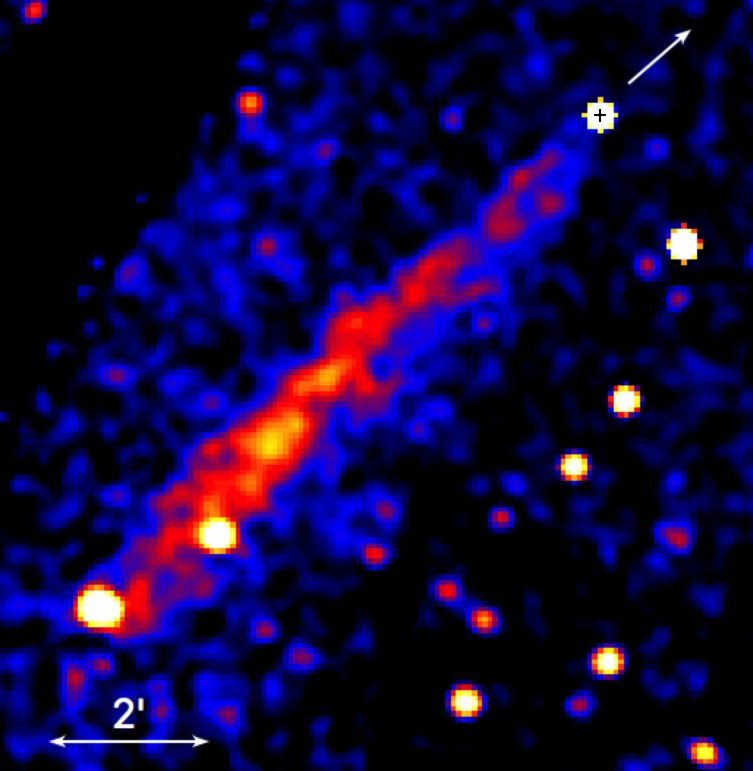}
\includegraphics[scale=0.24]{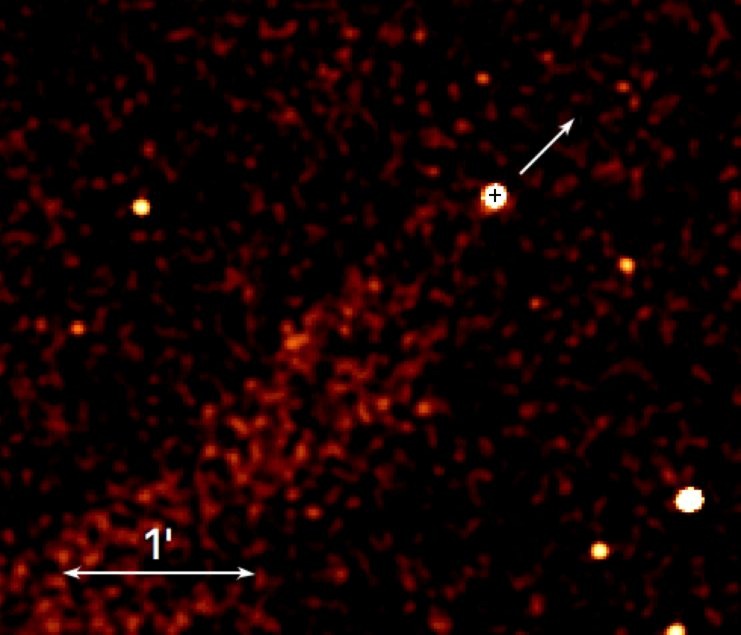}
\caption{Left: Merged {\sl Chandra} ACIS image of the tail behind PSR J0357+3205
(0.5--6 keV, 4 observations, 136 ks total exposure).
Right: Zoomed-in part of the same image showing the lack of tail emission
in the pulsar vicinity.}
\label{fig:J0357}
\end{figure}

An interesting X-ray nebula created by 
a radio-quiet $\gamma$-ray  pulsar J0357+3205
($P=444$ ms, $\tau_{\rm sd}=540$ kyr, $\dot{E}=5.9\times 10^{33}$ erg s$^{-1}$)
is shown in Figure \ref{fig:J0357} \citep{deluca2011}.
The pulsar's proper motion is $\mu=164\pm20$ mas yr$^{-1}$
\citep{deluca2013}, but the distance
is unknown. For an assumed $d=0.5$ kpc, its transverse velocity is
$V_\perp = (389\pm 47) d_{0.5}$ km s$^{-1}$.
The {\sl Chandra} and {\sl XMM-Newton} images show a $9'$ ($1.3 d_{0.5}$ pc) 
long, relatively straight tail
behind the pulsar, but no PWN head is seen. Moreover, the tail is detached 
from the pulsar (not seen up to $50''$),
 and its brightness increases with increasing distance from
the pulsar, reaching a maximum at 
about $4'$.
Another unusual property of the tail is the asymmetric brightness profile
across the tail, with a sharp northeastern edge, 
resembling the ``misaligned outflow'' in the
Gutar nebula.
The tail's luminosity, $L_{\rm 0.5-10\,keV}\approx 8.8\times 10^{30}d_{0.5}^2$ erg s$^{-1}$, is a fraction of $\sim 1.5\times 10^{-3}$ of the pulsar's spin-down power.

The tail's spectrum fits a PL model with a photon index $\Gamma\approx 2$,
without a significant dependence on the distance from the pulsar. Such a
spectrum is consistent with the synchrotron emission from relativistic
electrons in a shocked PW, but the lack of a PWN head is challenging
for such an interpretation. Another potential problem 
is the relatively low value for the maximum synchrotron photon energy,
$E_{\rm max} \lesssim 30\eta\sigma B_{-5}$ keV (see Equation
\ref{max-energy}), which is below 1 keV if $\eta\sigma B_{-5} \lesssim
0.03$.  To circumvent these problems, \citet{marelli2013} suggest that
the tail's emission is in fact thermal bremsstrahlung from the
shocked ISM material with a temperature of about 4 keV. To heat the
ISM up to such a high temperature, a very high pulsar velocity,
$V_{\rm psr}\sim 1900$ km s$^{-1}$, is required (see Equation
\ref{ion-temperature}), larger than observed for any other pulsar,
which would also imply a small angle, $i <20^\circ$, between the
velocity vector and the line of sight. The lack of diffuse emission at
small distances from the pulsar could then be caused by the considerable
time required for the energy transfer from ions, heated by the shock,
to radiating electrons.  If this scenario is confirmed by future
observations, the J0357+3205 nebula would be the first example of a
new class of thermally emitting nebulae associated with high-velocity
pulsars.

\subsubsection{Unexpectedly faint X-ray PWNe}
\label{subsubsec:faint_pwne}

From the above examples one could expect that any sufficiently powerful
pulsar that has left its parent SNR produces a head-tail PWN, possibly
with some misaligned outflows. However, observations of several nearby
pulsars show either very faint extended emission around the pulsar or
no extended emission at all. The most convincing examples of very faint
(or undetected) PWNe were provided by observations of nearby 
pulsars B1055--52 
and B0656+14.

PSR B1055--52 ($P=197$ ms, $\dot{E}=3.0\times 10^{34}$ erg s$^{-1}$,
 $\tau_{\rm sd}=535$ kyr) 
is a bright $\gamma$-ray pulsar at an estimated distance of
$\sim 350$ pc \citep{mignani2010}. Its proper motion, $\mu=42\pm5$ mas yr$^{-1}$, corresponds
to the tranverse pulsar velocity $V_\perp \approx 70 d_{0.35}$ km s$^{-1}$.
A dedicated 56 ks  {\sl Chandra} ACIS observation by \citet{posselt2015} 
showed 
some enhancement (with respect to the model PSF) in radial
count distribution from $2''$ to $20''$, better seen in the soft X-ray band
(0.3--1 keV), corresponding to the luminosity of 1--$2\times 10^{29} d_{0.35}^2$ erg s$^{-1}$, which is 
(3--$6)\times 10^{-6} d_{0.35}^3$ of the pulsar's
spindown power. The alleged extended emission showed only a hint of 
azimuthal asymmetry (an excess in the quadrant that includes the proper
motion direction), at a $3\sigma$ level. This extended emission (if real)
could be, at least partly, a dust scattering halo, but a very faint
X-ray PWN cannot be excluded. \citet{posselt2015} speculate that such a
faint, nearly round PWN could be produced if the pulsar is moving away from us
almost along the line
of sight, i.e., $V_{\rm psr}\gg V_{\perp}$. It, however, remains
unclear whether this interpretation is correct.

Very similar results were obtained by \citet{birzan2016} for PSR
B0656+14 ($P=385$ ms, $\dot{E}=3.8\times 10^{34}$ erg s$^{-1}$,
$\tau_{\rm sd} = 111$ kyr, and $d=0.29\pm 0.03$ kpc from parallax
measurements).  This pulsar also has a low transverse velocity,
$V_\perp=(60\pm 7)d_{0.29}$ km s$^{-1}$. From the analysis of archival
{\sl Chandra} ACIS and HRC data, \citet{birzan2016} found a slight
enhancement over the model PSF in an annulus of about $3''$--$15''$
around the pulsar, with a luminosity of $\sim 8\times 10^{28}
d_{0.29}^2$ erg s$^{-1}$. This luminosity is $\sim 2\times
10^{-6}d_{0.29}^2$ of the pulsar's spindown power, a factor of $\sim
7$ lower than the X-ray efficiency of the PWN of Geminga that has a
similar (slightly higher) spindown power and is a factor of 3 older
than PSR B0656+14. No azimuthal asymmetry was detected in the
images. The spectrum of the enhancement is apparently very soft,
$\Gamma\sim 8$, but its uncertainty is very large because the imaging
ACIS observation was very short, about 5 ks. As in the case of
B1055--52, the extended emission (if real) could be a combination of a
dust scattering halo and a PWN created by the pulsar moving almost
along the line of sight. The PWN and halo contributions could be
disentangled from a longer ACIS observation.

Thus, a plausible explanation for the lack of the expected head-tail
morphology and a very low PWN luminosity might be due to smallness of
the angle between the pulsar velocity direction and the line of sight,
which is also indicated by the small values of $V_\perp$. However, the
transverse velocity of PSR B0355+54 is similarly low, but that pulsar
is accompanied by a PWN with a rich structure (see Figure
\ref{fig:B0355}). Moreover, the spindown power of PSR B0355+54 is similar
to those of B1055--52 and B0656+14, but its PWN luminosity is at least
a factor of 300 higher. Obviously, there must be some other factors
that affect the X-ray efficiency and appearance of PWNe created by
pulsars moving in the ISM. A possible reason for these differences
could be different orientations of the pulsar rotational axes (hence
the equatorial planes) with respect to their velocities.  Another
parameter on which the PWN properties should depend is the angle
between the spin and magnetic axes, which affects the conversion of
the PW magnetic energy into kinetic energy and particle
acceleration.  In particular, the very low efficiencies of some PWNe
could be associated with nearly aligned spin and magnetic axes.  To
check these hypotheses, it would be useful to look for a correlation
between the PWN properties and multiwavelength pulsations.

\subsubsection{General overview of the X-ray PWNe created by supersonically 
moving pulsars}
\label{subsubsec:overview}

About 15 X-ray 
PWNe created by pulsars 
moving through the ISM
have been detected.
The spindown powers 
$\dot{E}$ of these pulsars are in the range from 
$1.2\times 10^{33}$ erg s$^{-1}$ (PSR B2224+65, the Guitar PWN) to 
$2.5\times 10^{36}$ erg s$^{-1}$ (PSR J1747--2958, the Mouse PWN).
Electrons/positrons of PWs of less powerful 
pulsars, which consitute the majority of 
rotation-powered pulsars, apparently cannot be accelerated to energies
high enough to emit X-ray synchrotron radiation, and even if the energy
is sufficient, the PWN luminosity may be too low to detect it, even from
nearby sources. 

The examples presented here show that most supersonically moving
pulsars are accompanied by tails, with typical lengths of a few
parsecs.  However, the appearances of PWN heads vary considerably in
different sources. Some of the well-resolved PWNe have a filled PWN
head morphology (e.g., the Mouse, B0355+54, J1741--2044) while others
show ``hollow'' morphologies (Geminga, J1509--5850). Moreover, there
is at least one example, J0357+3205, which shows a long tail but no
resolved PWN head around the pulsar.  The diversity of PWN heads
suggests that PWs of old pulsars are anisotropic, perhaps concentrated
around the equatorial plane (as in the Crab and some other young
pulsars) in many cases.  The different appearances of the compact PWN
components could be due to different orientations of the pulsar's spin
axis with respect to the velocity direction and the line of sight.  A
particularly puzzling morphology is seen in the Geminga PWN, with its
three ``tails'', which can be considered as a hollow-morphology
compact PWN component observed from a close distance (it might have a
much longer tail that is perhaps too faint to be detected by the
current instruments).

\begin{figure}
\includegraphics[scale=0.24]{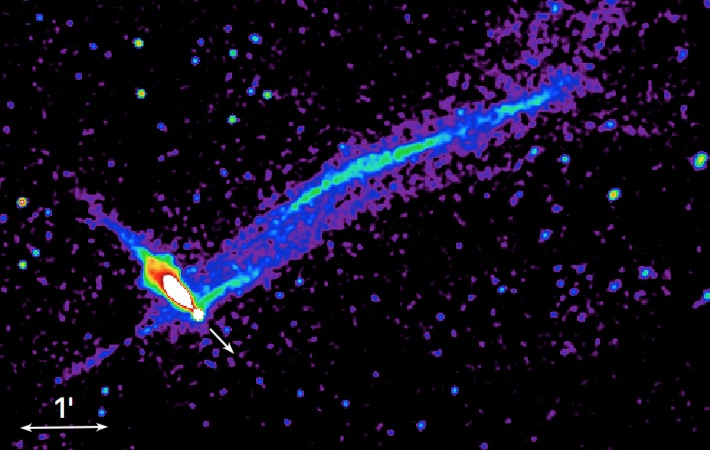}
\includegraphics[scale=0.45]{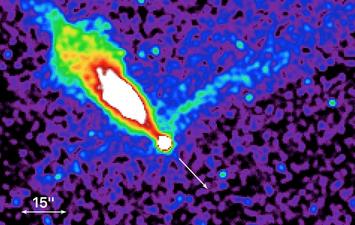}
\caption{{\sl Chandra} ACIS image of the Lighthouse PWN
(0.5--8 keV, 
300 ks) created by PSR J1101--6101 ($\dot{E}=1.4\times 10^{36}$ erg s$^{-1}$, $\tau_{\rm sd}=116$ kyr) moving from the SNR MSH\,11--61A \citep{pavan2016}.
The image shows a bright tail behind the fast-moving pulsar ($V_\perp\sim 1000 d_7$ km s$^{-1}$) and an $11d_7$ pc long jet-like feature, possibly another example
of a ``misaligned outflow''.
}
\label{fig:lighthouse}
\end{figure}

Quite unexpected was the discovery of ``misaligned outflows'' in X-ray
observations of several pulsars: B2224+65 (the Guitar), J1509--5850,
B0355+54, described above, and likely the spectacular Lighthouse
nebula created by PSR J1101--6101 (\citealt{pavan2016} and references
therein; see Figure \ref{fig:lighthouse}). Their nature still remains
puzzling.  A hypothesis was suggested by \citet{bandiera2008} that
these features are produced by synchrotron radiation of very high
energy particles leaked from the TSs into the ambient ISM along the
ISM magnetic field, but this interpretation remains to be confirmed by
quantitative modeling.

Although the flow speeds along the tails or misaligned outflows have
not been directly measured, they have been crudely estimated for some
tails based on circumstantial arguments. These estimates show that
flow speeds considerably exceed the pulsar speeds, but they are well
below the mildly relativistic speeds predicted by the numerical
simulations.

The X-ray efficiencies, $\eta_X = L_X/\dot{E}$, of the observed PWNe
vary between $\lesssim 2\times 10^{-6}$ (the alleged PWN around PSR
B0656+14) and $\sim 10^{-2}$ (the Mouse). The reason for such a huge
scatter remains unclear. At least partly, it can be due to different
orientations of the pulsar equatorial planes, where the PWs are
presumably concentrated, with respect to the pulsar velocities.  It
also may be that the fraction $\xi_w$ of the spindown power lost to
the PW is different in different pulsars (because the fraction of
$\dot{E}$ radiated from pulsar magnetospheres depends on pulsar
parameters).  Another likely reason for different PWN efficiencies is
associated with conversion of the magnetic PW energy into the kinetic
energy of particles, which should increase with increasing angle
between the magnetic and spin axes.

The X-ray spectra of PWNe created by supersonically moving pulsars are
usually well described by a PL model, which supports their synchrotron
interpretation.  Typical spectral slopes $\Gamma$ in the compact
nebula components (PWN heads) are in the range of 1.5--2.0, but the
lateral tails of the Geminga PWN are much harder, $\Gamma\sim
0.7$--1.0.  Some of the PWN tails (e.g., in the Mouse and Lighthouse)
show a substantial softening with increasing distance from the pulsar,
up to $\Delta\Gamma\approx 1.0$--1.5, while others show no softening
at all, sometimes even a hint of hardening. The fast softening is an
indication of a relatively high magnetic field (e.g., up to a few
hundreds of $\mu$G in the Mouse, the highest value found so far, which
is a factor of 10 higher than typical magnetic fields).

A few head-tail PWNe have been detected in the radio. The radio tails
are usually longer than the X-ray ones, as expected from the
synchrotron cooling. The additional radio data allow one to examine a
broad-band PWN spectrum, which is usually harder in the radio than in
the X-rays, and get an idea about the spectrum of emitting particles.
Measuring the spatially resolved radio polarization makes it possible
to map the directions of the magnetic field within the PWN, but the
two PWNe for which such mapping was done (the Mouse and J1509--5850)
show very different distributions.

The X-ray spectral slopes $\Gamma <2$ suggest that the main
contribution to the total synchrotron PWN luminosities are provided by
photons with energies above the observed X-ray range, i.e., above
$\sim 10$ keV.  However, none of the head-tail PWNe has been detected
at hard X-rays or $\gamma$-rays, perhaps because the current detectors
are not sensitive enough.

\subsection{Open questions}
\label{subsec:open_questions}

Although many of the observed properties of PWNe produced by
supesonically moving pulsars are qualitatively understood, there
remain several problems that require further investigations, both
observational and theoretical.

First of all, we should understand the reason(s) for the the {\em
  great diversity of PWN shapes}. Although elongated X-ray tails
behind the moving pulsars have been observed in many of them, it is
not quite clear which parameters determine the tail properties (shape,
length, collimation and divergence, separation from the pulsar in some
cases). Even less clear is the origin of the divesrity of PWN heads
(e.g., filled vs.\ hollow morphology).  One of such parameters is
obviously the angle between the velocity vector and the line of sight,
but other parameters, such as the angle between the spin and magnetic
axes, and between the spin axis and the pulsar velocity, can play an
important role. It is also possible that the direction and strength of
the ambient ISM magnetic field can affect the observed surface
brightness distribution.  To assess the contribution of the different
factors, a study of correlation of the PWN shape with the shapes and
phases of pulsar pulses at different wavelengths (e.g., radio and
$\gamma$-rays), supplemented by PWN modeling in the case of
anisotropic PW, would be particularly useful.

The most puzzling features among the recently discovered PWN
components are the ``{\em misaligned outflows}'', whose directions are
strongly misaligned with respect to the pulsar velocities. It is
tempting to interpret them as jets along the pulsar spin axes, similar
to those observed in PWNe of young pulsars, but such jets are expected
to be bent by the ram pressure on much smaller scales than the
observed lengths of these nearly straight, elongated features. Only
qualitative interpretations of such features have been suggested so
far, which remain to be confirmed by quantitative modeling.

It remains unclear whether ``ordinary'' {\em jets along the spin axes}
have been detected in old PWNe outside SNRs. May it be that the
outflows along the spin axis are less powerful (at least, less
luminous) in old pulsars than in young ones? To answer this question,
we should obtain independent information on the spin axis directions,
which could be done with the aid of multiwavelength polarimetry, in
addition to the pulse analysis.

Since the X-ray PWN emission is synchrotron radiation from
relativistic electrons and/or positrons, one should expect softening
of the PWN spectrum with increasing distance from the pulsar due to
synchrotron cooling.  Such softening has indeed been observed in
some of the tails, but other tails, as well as the misaligned
outflows, show {\em no spectral softening} at all, but hints of
hardening. What is the reason for such behavior? Is it
an indication of an additional (re)acceleration (heating) along the
tails? What is the acceleration mechanism? Is it the same mechanism
that is responsible for the unusually hard spectra of the Geminga's
lateral tail?  Why does it operate only in some tails?  To answer
these questions, deep high-resolution X-ray observations are required,
which would allow spatially resolved spectral analysis to accurately
measure the spectral changes. If the lack of softening (or even
spectral hardening) is confirmed, possible acceleration mechanisms
(e.g., the Fermi acceleration at fronts of oblique shocks or magnetic
turbulence) should be studied.

Another puzzling problem is the {\em very low X-ray efficiency (or
  even the absence) of PWNe around some nearby pulsars} that are
powerful enough to create an observable X-ray PWN.  While it might be
partly expained by closeness of the pulsar velocity direction to the
line of sight, it is certainly not a full explanation.  Are these PWNe
so faint because the conversion of the PW magnetic energy into the
particle energy (e.g., via magnetic field reconnection in the striped
wind zone) is inefficient, as expected for pulsars with nearly aligned
magnetic and spin axes?  Are there other mechanisms that suppress the
production and acceleration of the particle component of PWs? To
understand the true reason, more nearby pulsars should be observed,
and the absence or presence of PWNe should be confronted with the
observed pulsar properties.

To conclude, significant progress in our understanding of PWNe of
supersonically moving pulsars has been achieved, thanks to the high
resolution and sensitivity of the {\sl Chandra} and {\sl XMM-Newton}
observatories, but there remain a number of open problems that could be
resolved with further deep X-ray observations and more realistic
modeling.

%
%

\section{Gamma-ray observations of pulsar-wind nebulae}
\label{sec:gamma}

Gamma-ray astronomy has entered a golden age during the last decade
thanks to the latest generation of space telescopes in the High-Energy
(HE; 0.1 $<$ E $<$ 100 GeV) domain (\lat, \agile), and ground-based
instruments in the very-high energy (VHE; 0.1 $<$ E $<$ 100 TeV)
domain, in particular the Imaging Atmospheric Cherenkov Telescopes
(IACTs) such as \hess, VERITAS and MAGIC, featuring unprecedented
performance\footnote{In what follows, the angular resolutions are
  provided as the 68\% containment radii. The \lat~sensitivity is
  given at TS = 25 in 10 y, for a source at ($\ell$, $b$) = (0\d,
  30\d) with Pass 8 data ({\small
    \url{https://www.slac.stanford.edu/exp/glast/groups/canda/lat_Performance.htm}}). The
  \hess~sensitivity is provided at 5 $\sigma$ in 25 h, for a source
  near zenith ({\small
    \url{https://www.mpi-hd.mpg.de/hfm/HESS/pages/home/proposals/}}). See
  also the dashed lines in Figure \ref{fig:hessj1356}.}. \lat, with a large field-of-view
of 2.4~sr, an angular resolution varying from $\sim$ 5\d~at 100~MeV to
$\lesssim$ 0.1\d~above 50 GeV, and a point-source sensitivity at the
level of $\sim$~10$^{-12}$~erg~cm$^{-2}$~s$^{-1}$ in the GeV domain,
has revealed more than 3000 ({\it 3FGL}) sources in the 0.1-300 GeV
band \citep{ref:3fgl} and 360 ({\it 2FHL}) sources above 50 GeV
\citep{ref:2fhl}. IACTs have so far detected more than 170 VHE sources,
two third of which being located in the Galaxy \footnote{See
  {\small\url{http://tevcat.uchicago.edu/}}}. In particular, the
\hess~experiment, exhibiting a 5\d-wide field-of-view, an angular
resolution of $\sim$ 0.06\d~and a point-source sensitivity of a 
few~10$^{-12}$~erg~cm$^{-2}$~s$^{-1}$ in the TeV domain, has surveyed the
inner Galactic Plane during ten years, resulting in the detection of
more than 70 sources\footnote{From Abramowski et
  al. (H.E.S.S. Collaboration) 2016, in prep. See also
  {\small\url{https://www.mpi-hd.mpg.de/hfm/HESS/pages/home/som/2016/01/}}}.

\begin{figure}[!hbt]
\centering
\includegraphics[width=0.9\linewidth]{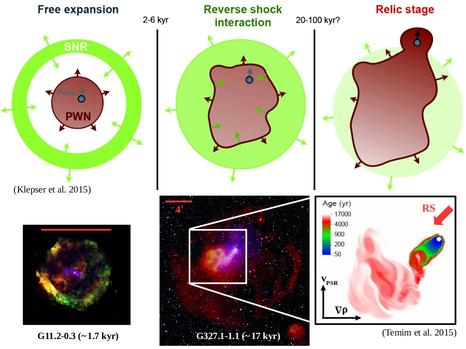}
\caption{PWN evolutionary phases. {\it Left:} young systems where the nebula lies at the center of the host shell-type SNR, as examplified by G11.2-0.3 seen in \xrays. {\it Middle:} evolved nebulae, after the so-called crushing phase caused by the interaction with the asymmetric reverse shock (RS) due to an inhomogeneous ISM and/or a high PSR's proper motion, as illustrated by G327.1$-$1.1 observed in radio (red) and \xrays~(blue). The inset image shows the age distribution of particles injected by the PSR some 17000 years ago, based on the hydrodynamical simulation of \cite{ref:temim15} with the assumed ISM density gradient and PSR's velocity depicted by the two arrows. {\it Right:} bow-shock nebulae, when the pulsar motion through the SNR or the ISM becomes supersonic.}
\label{fig:PWNevolution}
\end{figure}

These GeV-TeV instruments offer for the first time the possibility to spatially resolve large enough sources such as middle-aged pulsar wind-nebulae (PWNe) and shell-type supernova remnants (SNRs), and to study source spectra over more than five decades in energy. Spectro-morphological studies are of prime importance as they allow one to pinpoint the origin of \gammaray~emission from SNR systems in which the pulsar (PSR), the associated wind-nebula, and the host SNR shell, can contribute to the observed emission. In this regard, while \lat~has detected more than 200 PSRs in the HE domain\footnote{{\small\url{https://confluence.slac.stanford.edu/display/GLAMCOG/Public+List+of+LAT-Detected+Gamma-Ray+Pulsars}}}, IACTs, and \hess~in particular, have revealed more than 30 TeV PWNe and PWN candidates associated with energetic PSRs \citep{ref:klepser13,ref:kargaltsev13}, a dozen of which being also detected at (multi-)GeV energies \citep{ref:acero13,ref:2fhl}. Nevertheless, a large fraction of the HE/VHE sources in the Galaxy still remain unassociated due to the limitations of gamma-ray instruments to precisely characterize the source morphologies in most cases and to the difficulty for radio/\xray~telescopes to reveal structures at scales of the order of the typical VHE source sizes ($\sigma \sim$ 0.1\d-0.3\d).

Gamma-ray emission from PWNe is usually interpreted in the so-called leptonic scenario\footnote{The possibility that a hadronic component could carry a significant fraction of the energy content in PWNe has been investigated in several works (see \cite{ref:dipalma16} and references therein), but the observational evidence of the presence of accelerated ions in PWNe, through \eg~pion production in hadronic interactions and the subsequent emission of $\gamma$-rays and neutrinos, has been so far elusive.}, where the accelerated electron-positron pairs emit through inverse Compton (IC) scattering off the ambient low-energy photons, from the CMB and the interstellar radiation fields (ISRFs) made of infrared emission from dust and optical/UV starlight. In this framework, HE/VHE observations allow one to derive the spectral distribution and energy content of the high-energy particles (provided the ISRFs are known), to reveal their spatial distribution, and to set important constraints on the particle acceleration and transport mechanisms. Furthermore, these measurements, when combined with radio/\xray~observations of the synchrotron (SC) component, provide estimates of the magnetic field strength and distribution in these sources, as we shall see below.

\begin{figure}[!hbt]
\centering
\includegraphics[width=\linewidth]{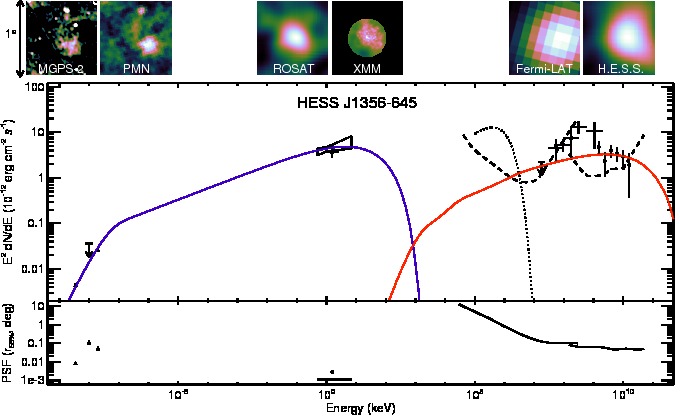}
\caption{Broadband spectrum of the PWN HESS~J1356$-$645 \citep{ref:hessj1356,ref:2fhl}. The blue and red solid curves give the SC and IC emission from a one-zone, time-independent, modeling performed in \cite{ref:hessj1356}. The dotted line represents the best-fit phase-averaged spectrum of PSR~J1357$-$6429 as obtained in \cite{ref:lemoine11}. The two dashed lines show the \lat~and \hess~point-source sensitivities (see text for more details). The 1\d-wide multi-wavelength images, as obtained in \cite{ref:hessj1356,ref:2fhl}, centered on the PSR position, are also shown. The lower panel presents the angular resolutions of the considered radio, X-ray and gamma-ray instruments.}
\label{fig:hessj1356}
\end{figure}

Two classes of $\gamma$-ray PWNe can be distinguished according to their evolutionary stages, as shown in Figure \ref{fig:PWNevolution} (no clear evidence of $\gamma$-ray emission associated with a bow-shock nebula [see section 3] has been reported so far). First, {\it young} PWNe, such as the Crab nebula, MSH~15$-${\it 5}2, G0.9+0.1, G21.5$-$0.9 and Kes~75 (discussed in section \ref{sec:gamma_youngPWNe}), are usually found at the center of their host shell-type SNRs and are seen as unresolved or compact $\gamma$-ray sources. HESS~J1818$-$154, a compact TeV source located at the center of the radio SNR G15.4+0.1, is worth to be mentioned as being the first PWN discovered by TeV observations in a composite SNR \citep{ref:hessj1818}. Second, {\it middle-aged} PWNe (with PSR characteristic ages of $\gtrsim$ 10$^{4}$ yr), such as HESS~J1356$-$645 (see Figure \ref{fig:hessj1356}), HESS~J1825$-$137 (section \ref{sec:hessj1825}), and the peculiar Vela~PWN (section \ref{sec:velax}), are usually resolved in the HE/VHE domains and are found to be significantly offset from the current position of the associated PSR, with large size ratios between the \xray~and $\gamma$-ray emission regions. The evolution of the SNR into an inhomogeneous ISM, through the interaction with an asymmetric reverse shock (RS), and/or the high PSR's velocity \citep{ref:blondin01,ref:vds04,ref:temim15} can lead to a displacement of the crushed PWN from the SNR center. Also, mean magnetic field strengths within these $\gamma$-ray PWNe as low as $\sim$ 3-5 $\mu$G are required in order to prevent the TeV-emitting electrons from suffering from severe radiative losses, enabling the majority of them to survive from (and hence probe) early epochs of the PWN evolution \citep{ref:djdj09}. All these effects can thus explain the large-offset X-ray-faint long-lived $\gamma$-ray sources as being the relic nebulae from the past history of the pulsar wind inside its host SNR.

\subsection{Young PWNe}
\label{sec:gamma_youngPWNe}

\begin{figure}[!hbt]
\centering
\includegraphics[width=\linewidth]{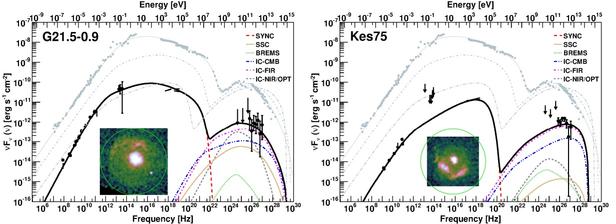}
\caption{Broadband spectra of G21.5-0.9 (left) and Kes~75 (right) with the best-fit models from \cite{ref:torres14} shown as solid lines. The grey dots and lines correspond to the multi-wavelength data and the associated models (at different ages) for the Crab nebula. \chan~\xray~images of these two PWNe with the typical \hess~PSF depicted as a green circle are also shown.}
\label{fig:g21_kes75}
\end{figure}

Several time-dependent models of the radiative (sometimes coupled with the dynamical) evolution of PWNe have been developed in order to estimate, from the observed broadband non-thermal emission, the energy distribution and content of particles and the wind magnetization, and to constrain the particle acceleration mechanisms at work in these sources \citep{ref:gelfand09,ref:fang10,ref:tanaka10,ref:tanaka11,ref:bucciantini11,ref:martin12,ref:vorster13,ref:torres14}. Figure \ref{fig:g21_kes75} shows the broadband spectra from G21.5-0.9 and Kes~75, two young PWNe respectively associated with the energetic pulsars PSR~J1833-1034 and PSR~J1846-0258 and detected in the VHE domain \citep{ref:djannati-atai08}, together with the best-fit models from \cite{ref:torres14}. The common findings of all the above-mentioned modelings are as follows: the particle energy distribution is well described by a broken power law with an intrinsic break\footnote{The modeling presented in \cite{ref:vorster13} assumes that the particle spectrum has a discontinuity at the transition between the low- and high-energy components.} (\ie~assumed to be of non-radiative origin) at a Lorentz factor of $\sim$ 10$^{5-6}$, and almost all of the young PWNe considered in these studies are particle-dominated. This is translated into large pair multiplicities ($\kappa \gtrsim$ 10$^4$) and magnetic fractions of $\lesssim$ a few percent (\ie~with magnetic fields lower than the equipartition values), with TeV emission dominated by IC scattering off (far-)infrared photons (with energy densities generally larger than the Galactic large-scale ISRFs provided by \cite{ref:porter05}). Although these constraints are valuable to pinpoint the particle acceleration mechanisms \citep{ref:amato15}, it should be noted that these modelings assume a one-zone emission region (in the 1D approximation, known to face several issues such as the so-called $\sigma$~problem, cf. a brief discussion in section 3.1), whereas the unresolved TeV emission from \eg~G21.5$-$0.9 and Kes~75 (see Figure \ref{fig:g21_kes75}) and the debated origin of the radio emission from these young PWNe could question this assumption. Also, relying on (or rescaling) the commonly used Galactic ISRFs as an estimate of the {\it local} fields from dust and stars is another major source of uncertainty which directly affects the predicted IC emission.

\subsection{Middle-aged PWNe}
\label{sec:gamma_evolvedPWNe}

As discussed above, middle-aged PWNe are observed in the HE/VHE domain as extended sources, offset from the PSR position. These two characteristics make them difficult to identify as such from $\gamma$-ray observations, and many of the so-called {\it dark} VHE sources, with no obvious radio/\xray~counterpart, could actually be such ancient nebulae \citep{ref:dejager09}. Besides dedicated pulsars search (with \lat~in particular) within the VHE source extent, multi-wavelength investigation, as shown in Figure \ref{fig:hessj1356} in the case of HESS~J1356$-$645, and energy-dependent $\gamma$-ray morphological analysis (as commonly done in \xrays) are the two means to unveil their nature.

\subsubsection{HESS~J1825$-$137: energy-dependent morphology and particle transport mechanisms}
\label{sec:hessj1825}

\begin{figure}[!hbt]
\centering
\includegraphics[height=5.8cm,valign=b]{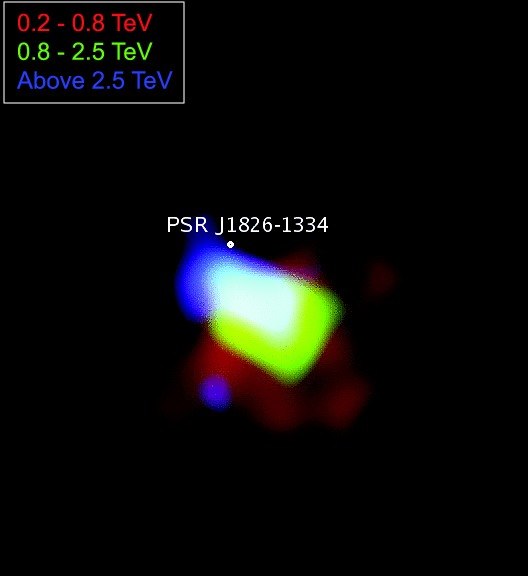}
\includegraphics[height=6.15cm,valign=b]{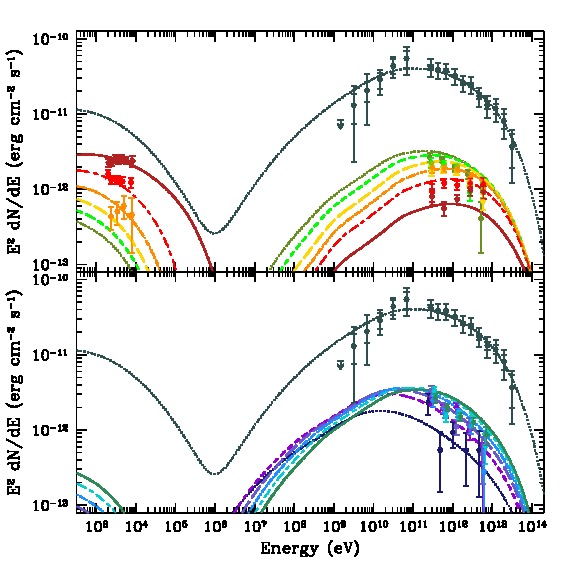}
\caption{{\it Left:} Three-color 3\d-large image of HESS~J1825-137 as observed with H.E.S.S. in different energy bands \citep{ref:funk08} showing the shrinking of the nebula with increasing energy. {\it Right:} Broadband spectra from the whole PWN (slate data points and dotted line) and from different wedges in HESS~J1825-137 with a color scale evolving from dark red (inner region) to green, blue and violet (outer region), as defined in \cite{ref:hessj1825}. Lines represent the best-fit model presented in \cite{ref:vanetten11} (see text for more details).}
\label{fig:hessj1825}
\end{figure}

HESS~J1825$-$137 is the archetypal middle-aged PWN, discovered by \hess~\citep{ref:hessj1825} as a bright, $\sim$1\d-large (\ie~$\sim$ 70 pc at 4 kpc) source offset from the energetic radio pulsar PSR~J1826$-$1334, and later detected with \lat~\citep{ref:grondin11}. A detailed spectro-morphological analysis has revealed for the first time in the VHE domain a steepening of the energy spectrum with increasing distance from the pulsar, likely due to the cooling losses of electrons during their transport in the nebula. This is illustrated by the three-colour image in Figure \ref{fig:hessj1825} (left) revealing the shrinking of the nebula with increasing energy and hence the physical connection between the TeV PWN with PSR~J1826-1334, also responsible for the compact hard-index \xray~nebula of size $\sim$ 30\asec~embedded in a $\sim$10\amin-large softer structure \citep{pavlov08,ref:vanetten11}. These $\gamma$-ray measurements, when combined with \xray~observations, provide important constraints on the particle injection, transport and cooling within the nebula: a detailed 3D time-dependent multi-zone spectro-morphological modeling \citep{ref:vanetten11} has shown a good agreement with the data by including radially decreasing advection velocity and magnetic field profiles and substantial particle diffusion in order to explain the presence of multi-TeV particles $\sim$ 80 pc away from the pulsar. Such a rapid diffusion is at odds with the toroidal magnetic field structure effective at smaller scales in many PWNe leading to strong magnetic confinement of particles.

\subsubsection{Vela~X: multi-wavelength picture and particle escape}
\label{sec:velax}

\begin{figure}[!hbt]
\centering
\includegraphics[height=4.5cm,valign=b]{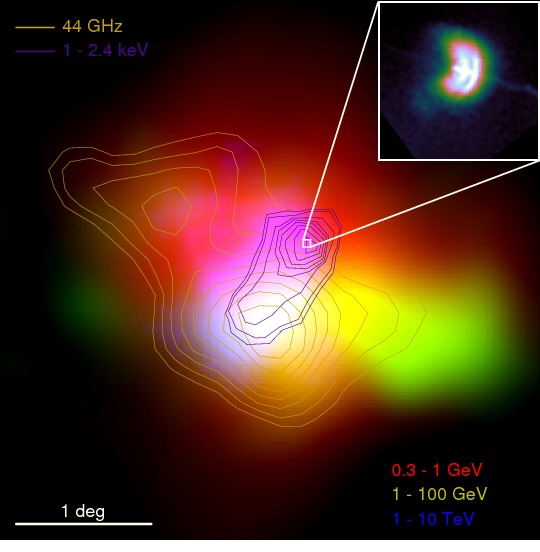}
\includegraphics[height=4cm,valign=b]{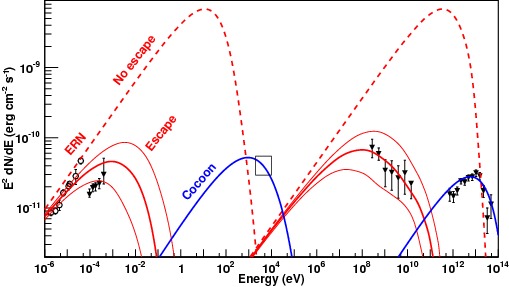}
\caption{{\it Left:} multi-wavelength picture of Vela PWN. The 4\d-wide three-color image (in Galactic coordinates) is composed of the 0.3-1 GeV (red) and 1-100 GeV (green) emission measured with \lat~\citep{ref:grondin13}, and of the 1-10 TeV (blue) emission detected with \hess~\citep{ref:velaX2}. Contours represent the 44 GHz radio emission measured by \planck~\citep{ref:planck16} and the X-ray emission above 1 keV observed with \rosat~(in blue). The well-known X-ray compact nebula around the Vela PSR is shown in the inset on the upper right corner (\chan/ACIS~0.3-10 keV image). {\it Right:} broadband spectrum of Vela PWN (ERN: Extended Radio Nebula) with (solid lines) and without (dashed lines) the assumption of particle escape throughout its evolution, according to \cite{ref:hinton11}.}
\label{fig:velaX}
\end{figure}

The energetic Vela pulsar (PSR~B0833$-$45, $\tau_c$ = 11 kyr, P = 89
ms, $\dot{E}$ = 7 $\times$ 10$^{36}$ erg s$^{-1}$), embedded in the
nearby 8\d-large Vela SNR located at d $\sim$ 290 pc, is known to
power several manifestations of wind nebulae seen at different spatial
scales: a compact \xray~nebula composed of two toroidal arcs at
sub-arcmin scales and a 4\amin~long collimated jet
\citep{ref:helfand01,pavlov03}, an extended hard \xray~emission (E
$>$ 18 keV) north of the pulsar \citep{ref:mattana11}, a
45\amin~elongated \xray~structure dubbed the {\it Vela cocoon},
also detected at TeV energies with \hess~\citep{ref:velaX1} and partially
coincident with a bright radio filament \citep{ref:lamassa08}, and the
Vela X nebula (referred to as the {\it Vela halo}, encompassing the
cocoon area), a large-scale 2\d~$\times$~3\d~non-thermal radio region
offset by $\sim$ 40\amin~from the pulsar that is also detected at GeV
energies with \lat~\citep{ref:abdo10}. Two distinct lepton populations
have been suggested to explain the different broadband spectra of the
latter two emission regions: a young population producing the
\xray/VHE cocoon and a relic one responsible for the radio/HE halo
\citep{ref:dejager09}, with similar magnetic field strengths of $\sim$
3-5 $\mu$G \citep{ref:abdo10}. However, such an interpretation has been
challenged by several new observational evidences: TeV emission beyond
the cocoon, extended over much of the halo and featuring a similar VHE
spectrum, has been detected with \hess~\citep{ref:velaX2}, and a
detailed spectro-morphological \lat~data analysis has revealed two new
spatial HE features matching the so-called northern and southern wings
of Vela~X as seen with \wmap~and \planck, with marginally different
spectra \citep{ref:grondin13}. Such a complex morphology, with several
emission components at different scales, is highlighted in the
multi-wavelength image shown in Figure \ref{fig:velaX} (left).

These new $\gamma$-ray measurements provide direct evidence that
high-energy leptons are present in the extended halo, $\sim$ 10 pc
away from the pulsar. In order to explain the steep spectra measured
with \lat, diffusive escape of particles from the radio nebula has
been invoked by \cite{ref:hinton11} (see Figure \ref{fig:velaX},
right), as in the case of HESS~J1825-137 discussed in section
\ref{sec:hessj1825}. While particle confinement is thought to be
important during the early PWN evolution, the interaction with the SNR
reverse shock could thus make the diffusion of particles out of the
nebula possible. Such an energy-dependent escape in this nearby PWN
should produce a clear signature in the local CR lepton spectrum
\citep{ref:hinton11,ref:dellatorre15}. Latest measurements with AMS-02
\citep{ref:aguilar14a,ref:aguilar14b} have shown an increase of the
positron fraction with increasing energy, which could be explained by
the Galactic PSR population and in particular by a few nearby PSRs
\citep{ref:hooper09,ref:dimauro14,ref:boudaud15}. Therefore, continuing
HE/VHE observations of middle-aged and bow-shock PWNe are crucial to
estimate the total energy content of high-energy particles residing
inside these sources in order to assess the importance of escape
mechanisms throughout their evolution, and hence their contribution to
the Galactic CR lepton spectra observed at Earth.

 


\section{Outflows in magnetars}
\label{sec:mag}

Magnetars differ from the rotation-powered NS discussed in the
previous sections because a strong magnetic field is the main energy
source which ultimately powers their persistent and bursting/flaring
emission \citep{mer15}.  Unlike ordinary pulsars, which show no or
very little variability, magnetars are characterized by a variety of
transient phenomena on timescales from a few milliseconds to years and
involving flux changes as large as several orders of magnitude.  There
are many indications that their magnetosphere is highly dynamic and
characterized by a complex non-dipolar geometry. It is thus expected
that magnetars might also be able to accelerate charged particles and
produce outflows which, in principle, can lead to the formation of
diffuse nebulae.

Owing to their long spin periods ($>$2 s), magnetars have rotational
energy losses much smaller than those of the energetic pulsars
typically associated with PWNe.  The magnetar with the highest spin-down
rate, 1E 1547.0--5408, has $\dot E_{ROT}=2.1\times10^{35}$ erg
s$^{-1}$, but more typical values are in the range
$\sim10^{32}-10^{34}$ erg s$^{-1}$.  While the ratio $L_{PWN} / \dot
E_{ROT}$ is a useful description of the efficiency of classical PWNe,
the analogous quantity for magnetars, $L_{MWN} / \dot E_{ROT}$ is less
relevant, because their nebulae might be powered, at least in part, by
magnetic energy.  In fact, the observation of time-variable radio and
hard X-ray emission after giant flares, as well as the presence of
long-lived nebulae around some magnetars, provide evidence that these
neutron stars can emit relativistic particle outflows.

In this section we discuss two kinds of observational results which
give evidence for relativistic outflows from magnetars: {\textit (a) }
the detection of time-variable emission associated to the occurrence
of giant flares, and {\textit (b)} the presence of persistent diffuse
emission at radio and/or X-ray energies around some magnetars.

\begin{figure}[b]
\centerline{ \includegraphics[scale=.45,angle=-90] {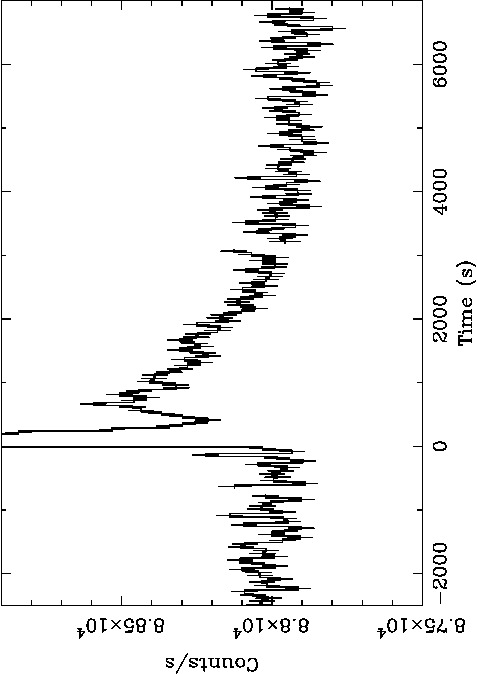}}
\caption{ High-energy ($>$80 keV) light curve of the 27 December 2004 giant flare of SGR 1806--20 measured with the INTEGRAL satellite (from \citet{mer05b}). After the bright flare truncated in the plot  (reaching $>2\times10^6$ counts s$^{-1}$) a long tail of hard X-ray emission, which peaks at t$\sim$500 s and lasts more  than one hour is clearly visible. }
\label{fig-sgr1806}      
\end{figure}

\subsection{Outflows during Giant Flares}
\label{ssec:mag-gf}

Giant flares are the most extreme manifestations of magnetars,
involving radiated energies up to about 10$^{46}$ erg (assuming
isotropic emission). They are rare events: only three have been
observed in more than 40 years (each one from a different source, in a
sample now totalling at least two dozen magnetars \citep{ola14}).  The
famous giant flare of 1979 March 5, from the Large Magellanic Cloud
source SGR 0526--66, was crucial for our understanding of magnetars,
but due to its large distance and unexpected occurrence, it could not
be studied in much detail.

More data could be obtained for the giant flare emitted on 1998 August
27 from the Galactic magnetar SGR 1900$+$14. Observations carried out
with the VLA about one week after the outburst revealed a faint
($\sim$0.3 mJy), unresolved radio source (angular diameter
$\theta<0.8''$), which became undetectable after a few days
\citep{fra99}. A power law index $\alpha=0.74\pm0.15$ was derived from
the fluxes at 1.43 and 4.86 GHz. This radio emission was interpreted
as a cloud of synchrotron emitting relativistic particles ejected
during the giant flare (or in the phase of intense bursting activity
which preceded it).  Simple equipartition arguments led to an estimate
for the nebula minimun energy of $1.6\times10^{43}~(d/10~{\rm
  kpc})^{17/7} ~(\theta/0.4'')^{9/7}$ erg.

The event of 2004 December 27 from SGR 1806--20 was by far the most
energetic and best studied giant flare, with an isotropic energy
release in hard X-rays and $\gamma$-rays of more than
$5\times10^{45}~(d/10~{\rm kpc})^2$ erg. The expanding radio nebula
detected after this flare could be observed for more than one year
\citep{gae05,cam05,gra06}. It had a peak flux of $\sim$170 mJy at 1.4
GHz in the first observation, carried out about one week after the
giant flare. The flux then decreased as a steep power law of time,
$F(t)\propto t^{-\delta}$ with $\delta\sim3$, and, after a brief
rebrightening at $\sim25$ days after the flare, it followed a
shallower power law decay with $\delta\sim1.1$.  The radio source was
spatially resolved and featured an elliptical shape; it expanded from
$\sim$60 mas to $\sim$400 mas in a couple of months, while at the same
time its centroid moved by about 200 mas, along the direction of the
elongation.  The power-law spectrum and linear polarization indicate
that the radio emission is synchrotron radiation.  The minimum energy
in the radio nebula was of the order of a few $10^{43}$ erg. This is
much larger than the energy available in the electron/positron pairs
escaping the initial fireball \citep{nak05}, implying that that the
relativistic flow powering the nebula was loaded by baryons or
Poynting flux.  Indeed, the observed properties of the expanding
nebula are well explained by an asymmetric ejection of few 10$^{24}$ g
of mildly relativistic baryons, $v/c\sim0.3~(d/10~{\rm kpc}) $, which
interact with a pre-existing shell of matter surrounding the magnetar
\citep{gel05,gra06}.

High-energy data obtained immediately after the SGR 1806--20 giant
flare provided additional evidence for the ejection of relativistic
matter. The $INTEGRAL$ satellite revealed emission at energy above
$\sim80$ keV, which peaked about 11 minutes after the start of the
flare and then decreased approximately as $F(t) \propto t^{-0.85}$
(Fig.~\ref{fig-sgr1806}). Evidence for this long-lasting high-energy
emission was later found also in the $Konus/WIND$ and $RHESSI$ data
\citep{fre07,bog07}, which indicated a duration of at least 10$^4$ s
and a hard spectrum (power-law photon index $\Gamma \sim 1.6$). No
pulsations at the NS rotation period of 7.56 s were seen in this
component, consistent with an origin far from the star surface and/or
magnetosphere.  This long-lasting emission can be interpreted as a
hard X-ray afterglow produced by the fireball ejected in the initial
spike of the giant flare \citep{mer05b}.

\subsection{Magnetar wind nebulae}
\label{ssec:mag-pwn}

\begin{figure}[b]
 \includegraphics[scale=.65]{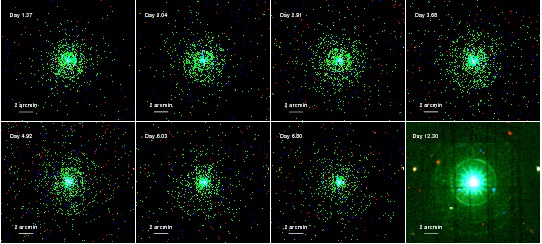}
\caption{Images of the expanding X-ray rings caused by interstellar
  dust scattering around the transient magnetar 1E 1547.0--5408 (from
  \citet{tie10b}).  The seven $Swift$ observations and the
  $XMM$-$Newton$ observation (bottom right panel) were obtained from
  one to 12 days after a very bright burst that occurred on 2009
  January 22. The apparent angular expansion of the three rings is due
  to the longer pathlength of the burst radiation scattered by three
  layers of dust along the line of sight. }
\label{fig-1e1547}      
\end{figure}

The identification and study of ``magnetar wind nebulae'' (MWNe) is
complicated by the fact that most magnetars lie in crowded and highly
absorbed regions of the Galactic plane and can be surrounded by
diffuse emission of different origins, such as supernova remnants,
molecular clouds, and H II regions.  For example, SGR 1806--20 was
initially associated with a variable radio nebula \citep{fra97}, but,
when a better localization of this magnetar was obtained, it became
apparent that a different object\footnote{the luminous blue variable
  star LBV 1806--20 \citep{fig04}.} is powering the radio
emission. Other magnetars are located inside radio-emitting supernova
remnants, but no signs of enhanced radio emission directly connected
with the neutron star has been found.  Therefore, there is no evidence
up to now for diffuse emission produced by magnetars in the radio
band, besides that of the transient nebulae associated to giant flares
described above.

In the X-ray range, where diffuse emission has been seen around
several magnetars, a further complication results from the effect of
interstellar dust scattering. Remarkable evidence for the importance
of this effect was demonstrated by the expanding X-ray rings
(Fig.~\ref{fig-1e1547}) seen around 1E 1547.0--5408 after its January
2009 outburst \citep{tie10}.

Excess X-ray emission over the $XMM$-$Newton$ PSF was detected at
radii from $\sim10''$ up to $\sim2'$ around the transient SGR
1833--0832 \citep{esp11}.  Its spectrum was softer than that of the
central source, as expected for a halo caused by interstellar dust,
due to the E$^{-2}$ dependence of the scattering cross section.  Given
the high absorption of SGR 1833--0832 ($N_H=10^{23}$ cm$^{-2}$), a
large amount of dust is likely present along its line of sight.  Thus,
the diffuse X-ray emission around this source, as well as that
reported for other highly absorbed sources like SGR 1900$+$14
\citep{kou01} and SGR 1806--20 \citep{kap02b,vig14}, is probably due
to dust scattering.  Recently, diffuse X-rays with a steep power-law
spectrum have been detected around SGR J1935$+$2154, on an angular
scale of about one arcmin with $XMM$-$Newton$
\citep{isr16,younes2017}. This could be a dust scattering halo, but a
possible contribution from a wind nebula cannot be excluded.


\begin{figure}[b]
\centerline{ \includegraphics[scale=.65] {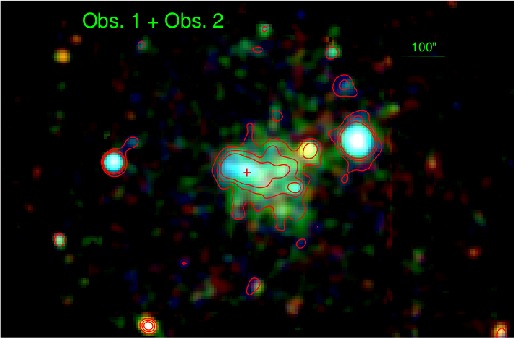}}
\caption{ X-ray image of the region of Swift J1834.9--0846 obtained
  with the $XMM$-$Newton$ satellite (from \citet{you16}).  The cross
  indicates the position of the magnetar. The colors code the X-ray
  energy (red 2--3 keV, green 3--4.5 keV, 4.5--10 keV blue). The
  contours are at the 2.5, 3.0, and 3.5 $\sigma$ level. For a distance
  of 4 kpc, 100$''$ correspond to $\sim$2 pc (horizontal bar).}
\label{fig-1e1834}      
\end{figure}

To date, the best evidence for a MWN is provided by the case of Swift
J1834.9--0846. This magnetar is surrounded by extended X-ray features
with different spatial scales (Fig.\ref{fig-1e1834}). The X-ray
emission within a radius of $\sim40''$, resolved in $Chandra$ images
\citep{kar12}, is most likely due to dust scattering of the outburst
emission of this transient. Its flux varied in correlation with that
of the central source, as expected for such small angles in case of
dust scattering \citep{esp13}. The more extended X-ray emission, with
an elongated shape ($\sim2'\times1'$) and a constant flux detected
with $XMM$-$Newton$ in 2005, 2011 and 2014, is instead best explained
as a MWN \citep{you12,you16}. In favour of this interpretation are the
relatively hard spectrum of the nebula, which is well fit by a power
law with photon index $\Gamma$=2.2, and the apparent lack of
variability. For a distance of 4 kpc, suggested by the possible
association of Swift J1834.9--0846 with the supernova remnant W41, the
nebula has an X-ray luminosity of $\sim3\times10^{33}$ erg s$^{-1}$.
This luminosity, which might be underestimated if the source is at a
larger distance, represents a considerable fraction of the the
spin-down power, $\dot E_{ROT}=2.1\times10^{34}$ erg s$^{-1}$, of
Swift J1834.9--0846.
   
We presented above the evidence for relativistic ejections during the
giant flares, but, according to the magnetar model, the acceleration
of particles in the magnetosphere is not restricted to these extremely
energetic events.  The normal bursting activity is expected to produce
particle outflows, and Alfv\'en waves can drive a steady wind also
during ``quiescent'' periods \citep{tho96,har99}. In this respect, it
is more meaningful to compare $L_{MWN}$ with the total reservoir of
magnetic energy. For example, if we assume for Swift J1834.9--0846 the
age of $\sim10^5$ yr estimated for the W41 supernova remnant
\citep{tia07}, the total energy radiated in X-rays in the nebula (a
few 10$^{38}$ erg) is a very small fraction of the magnetic energy
$\sim R^2 \Delta R\, \, B^2 \sim 10^{45} B^2_{14}$ erg, if the field
is in a crustal depth of $\Delta R \sim 1$ km (although probably only
a small part of such energy may be available for powering a nebula).

Indeed, several authors recently proposed models in which the nebula
of Swift J1834.9--0846 is magnetically-powered \citep{ton16,gra17}.
Considering the few observational data available up to now for this
alleged MWN, there is considerable degeneracy in the numerous involved
parameters and it is not surprising that such models led to rather
different estimates for the PWN and magnetar properties. For example,
the wind braking scenario of \citet{ton16} invokes a flow of particles
with luminosity between 10$^{36}$ and 10$^{38}$ erg s$^{-1}$ and
requires a rather high magnetic field in the nebula, $\sim10^{-4}$ G,
while \citet{gra17} estimate a likely upper limit of $\sim30\mu$G.  An
alternative interpretation, in which the Swift J1834.9--0846 nebula is
powered by rotational energy, like normal PWNe, has been proposed by
\citet{tor17}. This is energetically possible if one considers the
rotationally-powered wind injected over the whole lifetime of the
magnetar and the reverberation effect due to the location in a
particularly dense ambient medium.

   
\section{Gamma-ray binaries: pulsar winds interacting with a massive companion}

The last decade has revealed a new group of gamma-ray emitters,
composed of a fast-rotating pulsar and a massive star. The emission,
which peaks in the MeV band, arises from the shocked region between
the stellar wind and the pulsar wind. The binary interaction typically
takes place around one AU from the pulsar, about 5 orders of magnitude
closer than for pulsars interacting with the ISM.  Although only a
handful of these systems have been discovered, theoretical work
benefits from the well-constrained environment created by the binary
companion and the wealth of information provided by orbital
variability.  As such, gamma-ray binaries have opened a new window on
pulsar wind physics.  This subject is reviewed fully elsewhere.

\section{Conclusions}

The phenomenology of pulsar winds and their impact on their
environments is a rich one, documented from radio to TeV photon
energies.  These objects pose various problems of particle
acceleration and propagation, magnetic-field evolution, and
neutron-star physics.  Here we have summarized the spectra and
morphology (primarily at X-ray and gamma-ray energies) of PWNe still
in their natal shell supernova remnants, young PWNe which for some
reason lack shells, PWNe from much older pulsars, interacting directly
with the interstellar medium; and magnetars, with tantalizing but not
yet definitive evidence for magnetar wind nebulae (MWNe).

Young PWNe exhibit their pulsars at near-birth properties, generally
with high spin-down luminosities.  The PWNe interact with expanding
ejecta in SNR interiors, tending to produce fairly symmetric objects,
at least compared to morphologies observed for much older objects.
These objects radiate synchrotron radiation from radio through hard
X-rays, whose SED contains important information on particle
acceleration and propagation.  The flat radio spectrum of these PWNe
is still unexplained; the few anomalous PWNe with steep radio spectra
have very unusual properties, such as very large ratios of radio to
X-ray sizes.  The spectra steepen in the mm--IR region of the
spectrum, generally by larger amounts than can be accounted for by the
simplest models.  

\nust\ observations above 10 keV have provided important new
information on the X-ray properties of young PWNe.  PWN sizes decrease
with increasing photon energy; that energy-dependence encodes
information on electron propagation (advective, diffusive, or a
combination).  It is still not clear if intrinsic spectral structure
is required to explain the observations, or whether synchrotron losses
in inhomogeneous sources with some combination of advection and
diffusion can explain the observed size decrease and spectral
steepening with increasing distance from the
pulsar. \nust\ observations also show unexpected slight steepening of
X-ray spectra in the interiors of the Crab Nebula and G21.5$-$0.9.
This effect may be a clue to some feature of particle energization
near the pulsar-wind shock not yet explored.

The spatial resolution of \nust\ allows the determination of
anisotropic morphological changes with increasing photon energy that
support strong asymmetry in the pulsar outflows beyond the termination
shock.  In the Crab Nebula, the energy-dependence of the torus radius
is about what is predicted in the Kennel-Coroniti spherically
symmetric model, while the NW counterjet length drops much more
steeply with increasing energy, at a rate comparable to that seen in
both MSH 15-5{\sl 2} and G21.5$-$0.9.  This resemblance suggests a
commonality of origin not yet explained.  MHD models of young PWNe
will need to have more accurate treatment of the spatial and temporal
evolution of relativistic-particle distributions in order to confront
these data effectively.

Pulsars that have left their host SNRs move in the ISM with supersonic
velocities, which drastically changes PWN morphologies. In particular,
long tails, with a typical length of a few parsecs, form behind the
fast-moving pulsars.  These tails, collimated by the ram pressure of
the oncoming ISM, are fast outflows of the shocked PW emitting
synchrotron radiation from radio to hard X-rays. In addition, PWN
``heads'' in the pulsar vicinity are seen in deep high-resolution
X-ray images. The appearance of PWN heads, very different in different
objects, depends on the direction of the pulsar velocity with respect
to the line of sight, as well as on the inclinations of the velocity
vector and the magnetic axis to the spin axis. The observed shapes of
the heads differ significantly from the current PWN models. The main
reason for these differences is likely the unrealistic model
assumption that the unshocked PW is isotropic.

In addition to the PWN tails and heads, deep X-ray observations have
revealed ``misaligned outflows'' in some PWNe, at large angles to the
pulsar velocities. Such outflows are not predicted by the current PWN
models, and their true nature remains puzzling.

The X-ray spectra of the head-tail PWNe are usually well described by
power-law models, with typical photon indices $\Gamma\sim 1.5$--2.0 in
the pulsar vicinity. This implies power-law spectral energy
distributions of the X-ray emitting relativistic electrons/positrons,
$dN_e/d\gamma \propto \gamma^{-p}$, with $p\sim 2$--3.  Some PWNe show
spectral softening with increasing distance from the pulsar, with
$\Delta\Gamma \approx 1.0$--1.5, which can be due to synchrotron
cooling.  However, no softening is seen in other PWNe, which remains
unexplained.  Moreover, the lateral tails of the Geminga PWN have
unusually hard spectra, $\Gamma\sim 0.7$--1.0, without measurable
spectral softening, which might suggest an unusual acceleration
mechanism for the radiating electrons.

Typical equipartition magnetic fields in head-tail PWNe are $\sim
10$--100 $\mu$G. Interpreting the spectral softening observed in some
tails as caused by synchrotron cooling, one can crudely estimate flow
speeds $V_{\rm flow} \sim 10^4$ km s$^{-1}$, much faster than the
pulsar speed but much slower than the mildly relativistic speeds
predicted by some models.

Head-tail PWNe have been detected in X-rays only for sufficently
powerful pulsars, $\dot{E}\gtrsim 10^{33}$ erg s$^{-1}$, likely
because electrons cannot be accelerated to high enough energies in
less powerful pulsars.  Unexpectedly, the X-ray efficiency,
$\eta_X=L_X/\dot{E}$, of head-tail PWNe shows a huge scatter, up to 4
orders of magnitude. Moreover, several nearby pulsars with $\dot{E} >
10^{34}$ erg s$^{-1}$ show either very faint PWNe or no PWN at
all. This might be associated with small angles between the spin and
magnetic axes, but the true reason remains unclear. To explain this
and other puzzling properties of the X-ray PWNe created by
supersonically moving pulsars, more such objects should be observed in
X-rays and other spectral domains, with long exposures and high
spatial resolution, and new, more realistic models should be
developed.

Gamma-ray observations of PWNe have provided important insights on
several aspects such as the spatial and spectral distributions of the
high-energy particles, their total energy content and the wind
magnetization. From a theoretical point of view, 3D time-dependent
multi-zone spectro-morphological models [\eg~\cite{ref:vanetten11}]
are required in order to grasp the processes of particle acceleration,
transport and escape in these nebulae. From an observational point of
view, the identification of these HE/VHE PWNe requires extensive
multi-wavelength investigation and/or detailed spectro-morphological
studies, only feasible for bright and resolved sources. In most cases,
the association of a $\gamma$-ray source with counterparts at other
wavelengths is uncertain\footnote{As illustrated by HESS~J1640-645
  \cite{ref:hessj1640} coincident with the radio SNR G338.3-0.0
  hosting one compact and one extended \xray~source in its center
  which turned out to be an energetic pulsar (PSR~J1640-4631,
  \cite{ref:gotthelf14}) and its wind nebula.}. In this regard, thanks
to a factor $\sim$10 improvement in sensitivity above 100 GeV, with
substantially better angular and spectral resolutions and wider
field-of-view than the current IACTs, the Cherenkov Telescope Array
(CTA; \cite{ref:cta13}), currently in its pre-production phase, will
have the potential to reveal hundreds of sources through a uniform
Galactic Plane survey \citep{ref:dubus13}. This will undoubtedly
trigger detailed morphological and spectral investigations towards a
large number of PWNe and meaningful population studies
\cite{ref:dow13}.

\begin{acknowledgement}

  SPR thanks the \nust\ team for their outstanding work and permission
  to present their PWN results.  GGP, OK, and NK are grateful to
  Andrei Bykov and Maxim Lyutikov for helpful discussions, and to
  Blagoy Rangelov and Bettina Posselt for their help in data analysis.
  Support for their work was provided by the National Aeronautics and
  Space Administration through {\sl Chandra} Awards G03-14082 and
  G03-14057 issued by the {\sl Chandra} X-ray Observatory Center,
  which is operated by the Smithsonian Astrophysical Observatory for
  and on behalf of the National Aeronautics Space Administration under
  contract NAS8-03060. The work was also partly supported by NASA
  grant NNX08AD71G.
MR is grateful to Jamie Cohen and Marco Ajello for providing 
the \lat~2FHL image towards HESS~J1356-645 and to Marie-H\'el\`ene
Grondin for the \lat~images of Vela~X. 

All of us warmly thank the organizers
for the invitation to participate in this very stimulating ISSI
workshop on {\it Jets and Winds in Pulsar Wind Nebulae, Gamma-ray
  Bursts and Blazars: Physics of Extreme Energy Release}, and the
International Space Science Institute (ISSI) for support.  
\end{acknowledgement}

\bibliographystyle{aps-nameyear}      
\bibliography{pwntot3}   

\end{document}